\documentclass{article}

\usepackage{arxiv}

\usepackage[utf8]{inputenc} 
\usepackage[T1]{fontenc}    
\usepackage{hyperref}       
\usepackage{url}            
\usepackage{booktabs}       
\usepackage{amsfonts}       
\usepackage{nicefrac}       
\usepackage{microtype}      
\usepackage{lipsum}		
\usepackage{graphicx}
\usepackage{natbib}
\usepackage{doi}

\usepackage{amsmath}
\usepackage{xcolor}
\usepackage{subfigure}

\newcommand{\bs}{\boldsymbol{s}}
\newcommand{\br}{\boldsymbol{r}}
\newcommand{\bw}{\boldsymbol{w}}
\newcommand{\bc}{\boldsymbol{C}}
\newcommand{\bb}{\boldsymbol{B}}
\newcommand{\bF}{\boldsymbol{F}}
\newcommand{\bd}{\boldsymbol{d}}
\newcommand{\by}{\boldsymbol{y}}
\newcommand{\bX}{\boldsymbol{X}}
\newcommand{\nstar}[1]{N_{\boldsymbol{#1}}^*}
\newcommand{\Dstars}{\boldsymbol{D}_{N_{\bs}^*}}

\newcommand{\Dstar}[1]{\boldsymbol{D}_{N_{#1}^*}}

\newcommand{\bbeta}{\boldsymbol{\beta}}

\title{Clustering the Nearest Neighbor Gaussian Process}

\author{ \href{https://orcid.org/0000-0001-5984-5794}{\includegraphics[scale=0.06]{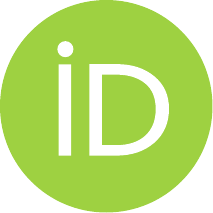}\hspace{1mm}Ashlynn~Crisp} \\
	Department of Mathematics and Statistics\\
	Portland State University\\
	Portland, OR \\
	\texttt{acrisp@pdx.edu} \\
	\And
	\href{https://orcid.org/0000-0002-2277-2912}{\includegraphics[scale=0.06]{orcid.pdf}\hspace{1mm}Andrew O.~Finley} \\
	Department of Statistics and Probability \\
	Michigan State University \\
        East Lansing, MI \\
	\texttt{finleya@msu.edu} \\
	\AND
    \href{https://orcid.org/0000-0002-2959-0281}{\includegraphics[scale=0.06]{orcid.pdf}\hspace{1mm}Daniel~ Taylor-Rodr\'{i}guez} \\
	Department of Mathematics and Statistics\\
	Portland State University\\
	Portland, OR \\
	\texttt{dantayrod@pdx.edu} 
}

\renewcommand{\shorttitle}{Clustering the Nearest Neighbor Gaussian Process}

\begin{document}

\maketitle

\begin{abstract}
    Gaussian processes are ubiquitous as the primary tool for modeling spatial data. However, the Gaussian process is limited by its $\mathcal{O}(n^3)$ cost, making direct parameter fitting algorithms infeasible for the scale of modern data collection initiatives. The Nearest Neighbor Gaussian Process (NNGP) was introduced as a scalable approximation to dense Gaussian processes which has been successful for $n\sim 10^6$ observations. This project introduces the \textit{clustered Nearest Neighbor Gaussian Process} (cNNGP) which reduces the computational and storage cost of the NNGP. The accuracy of parameter estimation and reduction in computational and memory storage requirements are demonstrated with simulated data, where the cNNGP provided comparable inference to that obtained with the NNGP, in a fraction of the sampling time. To showcase the method's performance, we  modeled biomass over the state of Maine using data collected by the Global Ecosystem Dynamics Investigation (GEDI) to generate wall-to-wall predictions over the state. In 16\% of the time, the cNNGP produced nearly indistinguishable inference and biomass prediction maps to those obtained with the NNGP. 
\end{abstract}


\section{Introduction}

Gaussian processes (GPs) have been the customary approach for modeling spatially dependent data. However, as modern datasets continue to grow increasingly large, fitting the original dense GP has become infeasible due to its $\mathcal{O}(n^3)$ computational complexity. As a consequence, several approaches have been developed to approximate spatial processes. 

Fixed rank ~\citep{cressie2008fixed}, lattice kriging ~\citep{nychka2015multiresolution} and stochastic partial differential equations ~\citep{lindgren2011explicit} assume the spatial process can be decomposed into a linear combination of basis functions. Spatial metakriging ~\citep{minsker2017robust} and spatial partitioning \citep[e.g.,][]{sang2011covariance} distribute computation by partitioning the data into subsets. Covariance tapering ~\citep{furrer2006covariance} assumes pairs of locations with a sufficiently small covariance are independent, improving computational performance by creating a sparse covariance matrix. ~\citet{banerjee2008gaussian} introduced predictive processes which use a set of ``knots'' as reference points to predict values at other locations. If the observations form a regular grid, periodic embedding ~\citep{guinness2019spectral} uses discrete Fourier transforms, providing computational efficiency through fast Fourier transforms. Gapfill ~\citep{gerber2018predicting} and the local approximate Gaussian process ~\citep{gramacy2015local} use local subsets to make predictions. 

Similarly to the Gapfill and local approximate GP methods, the Nearest Neighbor Gaussian Process (NNGP) ~\citep{datta2016hierarchical} reduces computational cost by only considering neighboring locations. In a series of competitions between many of these approximation strategies, ~\cite{heaton2019case} and \cite{hong2023competition} found that the NNGP had competitive computational and predictive performance. Additionally, ~\citet{finley2019efficient} provided alternative formulations of the NNGP to further improve computational efficiency and convergence, while ~\citet{guinness2018permutation} showed grouping calculations for observations can improve model accuracy. 

Another extension of the NNGP is the blockNNGP \citep{Quiroz_Prates_Dey_Rue_2023} which divides the domain into blocks and treats each block analogously to a location in the NNGP so that each block has neighbor blocks to condition the current block's spatial effects on. This strategy allows the model to better capture long range spatial correlation by using locations in neighbor blocks. The blockNNGP uses INLA to perform fast model fitting. Along similar lines, aiming to reduce the redundant evaluations and mitigate the computational burden of the traditional NNGP, \citet{Pan25} introduce the block Vecchia, where blocks of observations are identified using k-means and the univariate conditionals are replaced by the multivariate conditional distributions for the blocks of observations.

Following a slightly different approach, \cite{peruzzi2022highly} developed the Meshed Gaussian Process (MGP), which makes use of a directed acyclic graph over a tesselation of the spatial domain to induce a highly scalable dependence structure that enables investigating problems over very large spatial domains. Specifically, the graph uses the tesselation together with a relatively small set of \textit{reference} locations to induce the dependence over the entire spatial domain, and renders the approach computationally scalable by assuming that locations in \textit{non-reference} sets are independent, conditional on the reference sets. This structure enables carrying out large-scale operations in parallel within a Gibbs sampler. While the MGP constitutes an innovative and highly scalable approach, it has some limitations worth pointing out. Most importantly, the conditional independence assumption within and across partitions of non-reference sets can be restrictive and may lead to artifacts (e.g., boundary effects between partitions). Further, the quality of the model is strongly influenced by how the spatial domain is partitioned, making the approach sensitive to parameter tuning and potentially challenging to implement.

In this article we introduce the \textit{clustered NNGP} (cNNGP), which offers comparable predictive performance to the NNGP while significantly reducing computational cost. The approach is motivated by the insight that, under stationarity, once the parameters are fixed the spatial covariance function is determined entirely by the distances between locations. Making use of this fact and the sparse structure of the NNGP, the cNNGP saves computation and memory by identifying groups of locations with similar distance patterns among the sets of points consisting of each location and its neighbors, and as such can be clustered together to reduce the number of operations required. The most computationally expensive steps in MCMC algorithms fitting NNGP models is calculating the kriging weights and the conditional variances to sample spatial random effects. As $n$ grows to be in the order of millions of observations, the $\mathcal{O}(nm^3)$ cost of an NNGP becomes substantial. Our proposed algorithm uses the same MCMC algorithm as the NNGP, but reduces the $nm^3$ cost of the NNGP to $\kappa m^3$, where $\kappa (<< n)$ is the number of clusters. The cNNGP also reduces the storage requirements for model fitting. The number of required matrices in memory drops from $n$ to $\kappa$ so that the memory cost reduces from $\mathcal{O}(nm^2)$ to $\mathcal{O}(\kappa m^2)$.  

While conceptualized independently, our approach is similar in spirit to the \emph{caching idea} introduced in \cite{peruzzi2022highly} for the MGP, where by building the reference points on a regularly-spaced lattice, the covariances among reference points can be cached and reused. However, our approach enables sharing covariance information without requiring the locations to be built on a lattice and has a relatively straightforward implementation.

The code implementing the cNNGP, including the simulations presented below is available \href{https://github.com/Ashlynn-C/cNNGP}{here}. Our implementation of the cNNGP adapted source code from the \textsf{spNNGP} (version  1.0.0) \citep{spNNGPPackage} and the \textsf{leaderCluster} (version 1.5) \citep{leaderCluster} R packages, with custom modifications to incorporate the proposed methods. 

The remainder of this paper is organized as follows. In Section 2 we introduce the methods, first providing a brief overview of the NNGP, and then describing the cNNGP algorithm. In Section \ref{sec:sim exp}, we provide results from a relatively large simulation study on four scenarios that are small enough to be able to compare the original NNGP with the proposed algorithm's performance. Here, too, we compare cNNGP with two related GP approximation methods. Then, in Section \ref{sec:gedi} we use the proposed approach to model biomass estimates from the Global Ecosystem Dynamics Investigation (GEDI) ~\citep{dubayah2020global}. Lastly, in Section \ref{sec:discussion} we conclude with a brief discussion of the cNNGP's benefits and limitations. 

\section{Methods}

In this section, we detail the methods underlying our approach, beginning with a brief overview of Gaussian processes and their approximation through the NNGP, followed by a description of our novel modifications to the standard NNGP framework.

\subsection{Background and Notation}

A Gaussian process is defined as a stochastic process over a region $\mathcal{D}$ such that the joint distribution of any finite collection of observations taken from locations in $\mathcal{D}$ follow a multivariate normal distribution. Let $\bw(\bs)$ denote a $q$-variate spatial random effect at location $\bs \in \mathcal{D}$. Assuming a zero-centered GP,  $(\bw(\bs_1)',\bw(\bs_2)',\ldots,\bw(\bs_n)')' \sim N(\boldsymbol{0},\boldsymbol{C}_{\mathcal{S}}(\boldsymbol\theta))$, where $\mathcal{S} =\{ \bs_1,\ldots\bs_n\} \in \mathcal{D}$ and $\bc_{\mathcal{S}}(\boldsymbol\theta)$ is the $nq\times nq$ cross-covariance matrix with entries parameterized by $\boldsymbol{\theta}$. 

GPs offer convenient marginal and conditional distributions for each $\bw(\bs_i)$.  This is an important point to consider, in light of the fact that the joint distribution of an $nq$-dimensional vector $\bw_S$ can be written as the product of conditional densities
\begin{equation}
    p(\bw_S)=p(\bw(\bs_1))\;p(\bw(\bs_2)|\bw(\bs_1))\ldots p(\bw(\bs_n)|\bw(\bs_{n-1})\ldots \bw(\bs_1)).
  \label{eq:NNGP-joint}
\end{equation}

Evaluating this product requires conditioning on sets up to size $n-1$, which for large $n$ implies a prohibitively costly computational burden. To overcome this limitation, \cite{datta2016hierarchical} developed the NNGP, which induces sparsity in the GP by assuming conditional independence of locations given their nearest neighbors, therefore drastically reducing the size of these conditioning sets. Note that, the joint density expressed by Equation \ref{eq:NNGP-joint} as a product of conditional densities,  implies an ordering on the locations that defines the conditioning sets. Although this ordering has no relevance in the evaluation of the dense GP, as shown in 
\citet{guinness2018permutation}, the quality of the approximation provided by the NNGP is strongly influenced by how observations are ordered.

Here, we use the \textsf{order\_maxmin\_exact} function provided by the \textsf{GPvecchia} package in R (version 0.1.7) \citep{gpvecchia}. Once the points are ordered, the conditioning sets are replaced with smaller sets consisting of the nearest $m$ neighbors of each location $\bs$. Let $N(\bs_i)\subseteq\{\bs_1,\bs_2,\ldots,\bs_{i-1}\}$ be the set of $m$ nearest neighbors of $\bs_i$ and $\bw_{N(\bs_i)}$ be the vector resulting from stacking the vectors $\bw(\br)$ for all $\br$ in $N(\bs_i)$. Then, Equation \ref{eq:NNGP-joint} can be approximated with: 

\begin{equation}
  \tilde{p}(\bw_S) = \prod_{i=1}^n p(\bw(\bs_i)|\bw_{N(\bs_i)}).
  \label{eq:NNGP-mjoint}
\end{equation}

Let $\bc_{\bs,N(\bs)}$ denote the $q\times m$ covariance matrix between $\bw(\bs)$ and $\bw_{N(s)}$, $\bc_{N(\bs)}$ the $m\times m$ covariance matrix for $\bw_{N(\bs)}$, and $C_{\bs}$ the $q\times q$ covariance matrix for $\bw(\bs)$. Equation \eqref{eq:NNGP-mjoint} implies conditional independence across locations given their respective nearest neighbors, with the conditional density for $\bf{w}(\bs)|\bw_{N(\bs)}$ given by
\begin{equation}
  \bw(\bs)|\bw_{N(\bs)} \sim N(\bb_{\bs}\bw_{N(\bs)}, \bF_{\bs}),
  \label{eq:NNGP-marg}
\end{equation}
with $\bb_{\bs} = \boldsymbol{C}_{\bs,N(\bs)}\boldsymbol{C}_{N(\bs)}^{-1}$ and $\bF_{\bs} = \boldsymbol{C}_{\bs} -\boldsymbol{C}_{\bs,N(\bs)}\boldsymbol{C}_{N(\bs)}^{-1}\boldsymbol{C}_{N(\bs),\bs}$ with $\bs \in \mathcal{D}$ as derived in \cite{datta2016hierarchical}. Hence, the approximated joint density is
\begin{equation}
  \tilde{p}(\bw_S) = \prod_{i=1}^n N(\bw(\bs_i)|\bb_{\bs_i}\bw_{N(\bs_i)}, \bF_{\bs_i}), \; \text{with $\bs_i \in \mathcal{S}$}.
\end{equation}
Using this approximation to the Gaussian Process, the NNGP can model the spatial dependence through the regression model given by 
\begin{equation}
    \bf{y(\bs)} = \bf{X(\bs)'\boldsymbol{\beta} + Z(\bs)'w(\bs) + \boldsymbol{\epsilon(\bs)}},
    \label{eq:model}
\end{equation}
where $\bf{\bs} \in \mathcal{D}$, $\bf{y(\bs)}$ is the $l$-variate response. Letting $p=\sum_{k=1}^l p_k$, $\boldsymbol{\beta}$ is the $p$-dimensional vector of regression coefficients for the fixed $l \times p$ block-diagonal matrix of  spatially referenced predictors $\bf{X(\bs)}'$, where the $k$th block corresponds to the $1 \times p_k$ vector $\boldsymbol{x}_k(\bs)'$ for the $k$th response.  Additionally, $\bf{Z(\bs)}'$ is the $l \times q$ design matrix for the spatial process $\bw(\bf{\bs})$, and $\boldsymbol{\epsilon}(\bs) \overset{\text{iid}}{\sim} N(\boldsymbol{0},\text{diag}(\tau^2_1,\ldots,\tau^2_l))$ denotes the $l \times 1$ measurement error vector, with $\{\tau_k^2>0: k=1,\ldots,\ell\}$. Even though \eqref{eq:model} is specified for a Gaussian response, extending the approach for non-Gaussian responses is easy to accommodate.

\subsection{Clustered Nearest Neighbor Gaussian Process} \label{sec:ccngp}

As mentioned earlier, the proposed strategy is motivated by the insight that, given parameters values, the NNGP covariance at any location solely depends on the distances between the location and its neighbors and the distances among those neighbors themselves. As such, the clustered NNGP speeds up computation by identifying approximately recurring patterns in these sets of distances and reducing the number of computations by exploiting these patterns. 

For ease of exposition, in what follows we assume that $l=q=1$. It is worth noting that for a location $\bs\in \mathcal{S}$ the prior density of $\bw(\bs)$ has mean $\bb_{\bs}\bw_{N(\bs)}$ and variance $\bF_{\bs}$, where $\bb_{\bs}$ and variance $\bF_{\bs}$ exclusively depend on $\boldsymbol{C}_{\bs,N(\bs)}$, $\boldsymbol{C}_{N(\bs)}$ and $\boldsymbol{C}_{\bs}$, all submatrices in the covariance of the $(m+1)$-dimensional vector $(\bw(\bs),\bw_{N(\bs)}')'$. Given the parameters, this covariance matrix is then determined by the distances between $\bs$ and $N(\bs)$, and between the neighbors themselves. As such, under the NNGP, if two locations have similar distances to their neighbor sets and similar pairwise distances among their neighbor sets, then their covariance to and among their neighbor sets will also be approximately the same. 

More formally, let $N_{\bf s}^* = \{\bf{s}\} \; \cup $ $N(\bf{s})$ and let $\Dstar{s}$ denote the ($m$ + 1) $\times$ ($m$ + 1) distance matrix among locations in $N_{\bf s} ^*$. Because $\bb_{\bs}$ and $\bF_{\bs}$ are determined solely by $\Dstar{s}$ and the covariance parameter vector $\boldsymbol{\theta}$, then for a particular $\boldsymbol{\theta}$, if two locations $\boldsymbol{s, r} \in \mathcal{S}$ have distance matrices such that $\Dstars$ = $\Dstar{r}$, then the covariance matrices $\boldsymbol{C}_{N_{\bs}^*}$ and $\boldsymbol{C}_{N_{\boldsymbol{r}}^*}$ among locations in $N_{\bf s}^*$ and in $N_{\bf r}^*$, respectively, are such that: 
\begin{equation*}
\boldsymbol{C}_{N_{\bs}^*} = \boldsymbol{C}_{N_{\boldsymbol{r}}^*} \Rightarrow
\begin{cases}    
  \bb_{\bs} = \bb_{\br}\\
   \bF_{\bs} = \bF_{\br}
\end{cases}.
\end{equation*}
By grouping together locations whose distance matrices $\Dstar{s}$ are similar, we can exploit these redundancies to dramatically reduce both computational cost and memory usage. This is the central idea behind the cNNGP.
 
The cNNGP begins with a pre-processing step to find the clusters of locations whose neighbor-distance matrices are similar. For each location $\bs$, define $\boldsymbol{d_s}$ the vector containing the strict lower triangular elements of $\Dstars$. Therefore, $\boldsymbol{d_s}$ is the $\binom{m+1}{2}$-dimensional vector of all pairwise distances among the locations in $\nstar{s} = \{\bs\} \cup N(\bs)$. Because neighbors in $\nstar{s}$ are ordered from closest to farthest neighbor, all $\boldsymbol{d}_s$ vectors share a consistent ordering.  

The collection $\{\boldsymbol{d}_s: \bs\in\mathcal{S}\}$ is then partitioned into groups using the chosen clustering algorithm. The first $m$ locations are excluded from the clustering since they have fewer than $m$ neighbors. Instead each of the first $m$ locations are assigned their own cluster. Thus, the clustering is applied only to locations $\{\bs_{m+1}, \bs_{m+2},\ldots,\bs_{n}\}$. We define the map between a location and its corresponding cluster label as $\nu: \mathcal{S}\rightarrow \{1,\dots,m,m+1,
\dots, m+\kappa\}$ where $\nu$ is an identity map for the first $m$ points such that $\nu(\bs) \in \{1,\dots,m\}$ for  $\bs \in \{1,\dots,m\}$ and $\nu(\bs) \in \{m+1,\dots,m+\kappa\}$ for all $\bs \in\{\bs_{m+1}, \bs_{m+2},\ldots,\bs_{n}\}$.

 Once the clustering step is complete, we use a representative distance matrices for each cluster and denote that distance matrix by
 $\bar{\boldsymbol{D}}_\ell$ for cluster $\ell$. Additionally, we denote by $\bar{\boldsymbol{C}}_{\ell}$ the $(m+1)\times(m+1)$ covariance matrix obtained from $\bar{\boldsymbol{D}}_\ell$ and the covariance parameters $\boldsymbol{\theta}$. The matrices $\bar{\boldsymbol{D}}_\ell$ remain fixed throughout the Markov chain Monte Carlo (MCMC) algorithm, while $\bar{\boldsymbol{C}}_{\ell}$ is updated at every iteration as $\boldsymbol{\theta}$ is resampled.
 
For any matrix $\boldsymbol{M}$, let $\boldsymbol{M}[\text{rows},\text{columns}]$ denote subsetting. Then, for each cluster $\ell$ compute the cluster's kriging weights $\bar{\boldsymbol{B}}_\ell$ and conditional variances $\bar{\boldsymbol{F}}_\ell$ as
\begin{align*}
    \bar{\boldsymbol{B}}_\ell &= \bar{\boldsymbol{C}}_{\ell}[1,2:(m+1)]\left(\bar{\boldsymbol{C}}_{\ell}[2:(m+1),2:(m+1)]\right)^{-1},\\
    \bar{\boldsymbol{F}}_\ell &= \bar{\boldsymbol{C}}_{\ell}[1,1]- \boldsymbol{B}_\ell\bar{\boldsymbol{C}}_{\ell}[2:(m+1),1].
\end{align*}
These are used to replace the location-specific weights and variances in the cNNGP approximation of the joint density $\tilde{p}(\bw_S) $, given by
\begin{equation}
  \tilde{p}(\bw_S) = \prod_{i=1}^n N(\bw(\bs_i)|\bar{\boldsymbol{B}}_{\nu(\bs_i)}\bw_{N(\bs_i)}, \bar{\bF}_{\nu(\bs_i)}).
  \label{eq:cNNGP-mjoint}
\end{equation}
Recall that the sparsity pattern of the NNGP precision matrix is determined solely by the directed nearest-neighbor graph. Given that the cNNGP retains the same neighbor sets, the underlying graph (and thus the sparsity structure of the precision) remains unchanged. The only modification introduced by the cNNGP is the substitution of location-specific kriging weights with cluster-level approximations; the pattern of nonzero entries dictated by the neighbor relationships is preserved exactly.

The MCMC implementation for the cNNGP model uses Gibbs steps for $\bbeta$, the set $\{\tau_k\}_{k=1}^{\ell}$, and $\{\bw(\bs):\bs\in\mathcal{D}\}$, and uses a Metropolis step for the spatial covariance parameters $\boldsymbol{\theta}$. For more details on the MCMC algorithm used throughout, we refer the reader Section 3 of \citet{datta2016hierarchical}. The cNNGP adaptation of this algorithm only requires substituting inside of the posterior distributions the location specific kriging weights $\bb_{\bs}$ and conditional variances $\bF_{\bs}$ for their corresponding cluster representatives $\bar{\bb}_{\nu(\bs)}$ and $\bar{\bF}_{\nu(\bs)}$.

Lastly, prediction is carried out as is customary, with predictions $\by^*(\boldsymbol{t})$ for a location $\boldsymbol{t}\in\mathcal{D}$ obtained from the posterior predictive distribution 
$$\by^*(\boldsymbol{t})|\{\by(\bs)\}_{\bs\in\mathcal{S}}\sim N(\bf{X}(\boldsymbol{t})'\boldsymbol{\beta} + Z(\bf{t})'w(\boldsymbol{t}) , \boldsymbol{\Psi}),$$ 
where $\boldsymbol{\Psi}= \text{diag}(\tau^2_1,\dots,\tau^2_l)$. If $\boldsymbol{t}\in\mathcal{D}\setminus\mathcal{S}$, then samples of $\bw^*(\boldsymbol{t})$ must be obtained from $N(\bf{B_t}\bw_{\mathcal{S}},\bf{F_t})$ first, and then posterior draws of $\by^*(\boldsymbol{t})$ are obtained from $N(\bf{X}(\boldsymbol{t})'\boldsymbol{\beta} + Z(\boldsymbol{t})'\bw^*(\boldsymbol{t}) , \bf{\Psi})$.

\subsection{Implementation Details}

An essential choice in the implementation of the cNNGP is the clustering strategy. One suitable alternative is Hartigan's leader algorithm ~\citep{10.5555/540298}, which provides an upper bound for the approximation error of $\Dstars$ and easily scales to a massive number of observations. 

The algorithm takes a radius $r$ and iterates through each $\boldsymbol{d}_{\bs_i}$ for $i = m+1,\ldots,n$ as follows. First, $\bd_{\bs_{m+1}}$ is set as the cluster leader for cluster 1. If $||\bd_{\bs_{m+2}} - \bd_{\bs_{m+1}} || > r$, then $\bd_{\bs_{m+2}}$ becomes the cluster leader for cluster 2, else it joins cluster 1. Similarly, the distance between $\bd_{\bs_{m+3}}$ and the previous clusters is computed. If $\bd_{\bs_{m+3}}$ is within the radius of a previous cluster, it is assigned to that cluster, otherwise it becomes a cluster leader. This continues for all $i = m+4, \ldots n$, resulting in $\kappa$ clusters of distance matrices where every matrix is within distance $r$ to its cluster leader.

To choose the number of clusters $\kappa$, we consider different values of $r$ and for each we run the clustering algorithm to obtain the $\kappa$ value corresponding to each $r$. We then find the elbow of the curve for the number of clusters ($\kappa$) vs. the radius ($r$). We recommend using some dimension reduction technique (e.g., principal components) of the $\binom{m+1}{2}$ columns of the matrix of neighbor set distances given by $(\bd_{\bs_1},\ldots, \bd_{\bs_n})'$ in the clustering to reduce the dimensionality of the vectors being clustered. For large datasets, we additionally suggest subsampling the PCA-reduced version of the distance vectors $\bd_{\bs}$ to identify the cluster leaders. In Section \ref{sec:gedi} we provide more details about how this process was carried out with the GEDI data set.

Code for the NNGP model is available through the \textsf{spNNGP} R package \citep{spnngp}, which provides many highly efficient routines spatial model fitting by leveraging parallel processing and highly efficient computational algorithms described in \citet{finley2019efficient}. To take advantage of the computational efficiencies built into the package, we implemented the proposed cNNGP model making the necessary modifications to the MCMC-based \emph{sequential} algorithm using the functions within \textsf{spNNGP}. The adapted code can be found in \href{https://github.com/Ashlynn-C/cNNGP}{this GitHub repository}. For our clustering implementation, we only made a minor modification to the the leader algorithm in the R package \citep{leaderCluster} so that it would produce both the cluster centroids and the cluster labels. 

\section{Simulation Experiments} \label{sec:sim exp}

We compare the performance of the cNNGP--in addition to that of the NNGP--to two other recently proposed strategies that aim to reduce the computational burden of fitting stationary spatial models. Specifically, here we consider the cNNGP and the NNGP with $m=10,20,30$ neighbors, the MGP \citep{peruzzi2022highly}, and the blockNNGP \citep{Quiroz_Prates_Dey_Rue_2023}. Even though most of these approaches can take advantage of multithreading, to make valid comparisons in terms of computational efficiency, each dataset-model combination was run single core on the same machine (Ryzen threadripper 3970x 32-core/64-thread processor and 256 GB RAM). All models considered use MCMC for estimation, except for the blockNNGP which uses INLA. With every MCMC-based model, 50,000 MCMC draws were sampled with a 30,0000 burn-in. We compare the models considered in terms of their computational efficiency, how well they recover the true model parameters, and in terms of their predictive ability.

\subsection{Data Generation}

Four different scenarios were generated by varying the sample size ($n= 2,500$ and $n = 10,000$ locations) and the decay parameter of the spatial covariance function (providing a short- and long-range scenario). Locations were drawn uniformly from a unit square, and the univariate response was sampled from the model $\by(\bs)={\bf x}'(\bs)\bbeta + \bw(\bs)+\epsilon(\bs)$, where $\bw(\cdot)\sim GP(0,C(\cdot,\cdot;\phi, \sigma^2))$, with exponential covariance function $C(\boldsymbol{s}_i,\boldsymbol{s}_j;\phi, \sigma^2) = \sigma^2 e^{-\phi ||\boldsymbol{s}_i-\boldsymbol{s}_j||},$ where the scale $\sigma^2=1$ and the spatial decay parameter $\phi$ takes values 11.51 and 2.88, resulting in effective spatial ranges of 0.2 and 0.8 distance units (where the effective spatial range is the distance at which the correlation drops to 0.1). Additionally, the true regression coefficient vector is set to $\boldsymbol{\beta}=(1,5)'$ and $\epsilon(\bs)\overset{iid}{\sim} N(0, \tau^2)$ with $\tau^2= 0.1$. 

For each scenario (a specific combination of values for $n$ and $\phi$) we sampled 30 different datasets (locations, predictions and responses), each of which was split into training and testing subsets using the \textsf{blockCV} \citep{blockCVPackage} R package (version 3.2.0) which produces  spatially separated folds. To do so, we placed a 20 by 20 grid of hexagons over the domain of the locations and divided each dataset into 5 folds with the function \textsf{cv\_spatial}. One fold was left as holdout for assessing prediction performance and the other four were used for model fitting, resulting in an approximately 80/20 split for fitting and testing. 

The priors for the parameters in all of the attempted models were assumed to be $\phi\sim\text{Uniform}(1,30)$, $\sigma^2\sim\text{InvGamma}(2,1)$, $\tau^2\sim\text{InvGamma}(2,0.1)$, and a flat prior for $\boldsymbol{\beta}$. For all MCMC models considered the regression coefficients were initialized at their OLS estimates $\hat{\boldsymbol{\beta}}_{OLS}$, $\phi$ was initialized at the mean of its prior distribution, and $\sigma^2$ and $\tau^2$ were each initialized at $\frac{1}{2} \text{Var}(\by-\bX\hat{\boldsymbol{\beta}}_{OLS})$, where $\by$ is the vector of responses and $\bX$ is the corresponding design matrix.

\subsection{Model Settings}

For the cNNGP, the values for the clustering radius $r$ were chosen by generating test coordinates for each dataset size and number of neighbors $m$. The cNNGP was run for each data scenario to decide on the number of MCMC iterations. Traceplots for these test datasets indicated that 50,000 iterations was more than sufficient to achieve convergence with the first 30,000 iterations as burn-in. The MGP was also fit on a test dataset to ensure that having 50,000 MCMC draws under all scenarios considered was sufficient.

The settings for the blockNNGP were chosen using as a reference the settings that performed well in the simulation results found in the manuscript \citep{Quiroz_Prates_Dey_Rue_2023}, where a very similar experiment setup to ours was carried out. Specifically, we chose a regular grid with $M = 36$ blocks and $nb = 2$ neighbor blocks. The code for the blockNNGP was obtained from \href{https://github.com/Zaidajqc/blockNNGP}{the GitHub repository} with the code used in \citet{Quiroz_Prates_Dey_Rue_2023} and we supplemented it with the R code needed to carry out predictions.

The settings for the MGP were chosen with input from the lead author in \citet{peruzzi2022highly} to allow the methods to perform as well as possible. The MGP models were fit using the \textsf{meshed} package (version 0.2.3) on CRAN \citep{meshedPackage}. In the package it is recommended to predict responses at new locations during model fitting since it is generally faster, however, to guarantee a fair comparison on model fitting times across methods, we used the \textsf{predict} function in the package after model fitting. In the MGP functions we set \textsf{block\_size}=30, and \textsf{forced\_grid=FALSE}. We made one minor edit to the \textsf{meshed} package to have the predict function return the draws for the spatial effects $\bw$, which are used eventually to calculate the Watanabe–Akaike information criterion.

\subsection{Simulation Results}

The simulation results exhibit a consistent pattern across all metrics. The cNNGP provides substantial computational savings and maintains excellent predictive performance, while showing bias in certain covariance parameters, especially under short–range dependence. These behaviors emerge naturally from the cNNGP construction, which clusters locations by similarity of their distance matrices $\Dstar{s}$ and replaces each site’s true distance matrix with its cluster representative $\bar{\boldsymbol{D}}_{\nu(\bs)}$.

\paragraph{Computational Efficiency}

\begin{figure}[ht]
    \centering
    \includegraphics[width=0.9\textwidth]{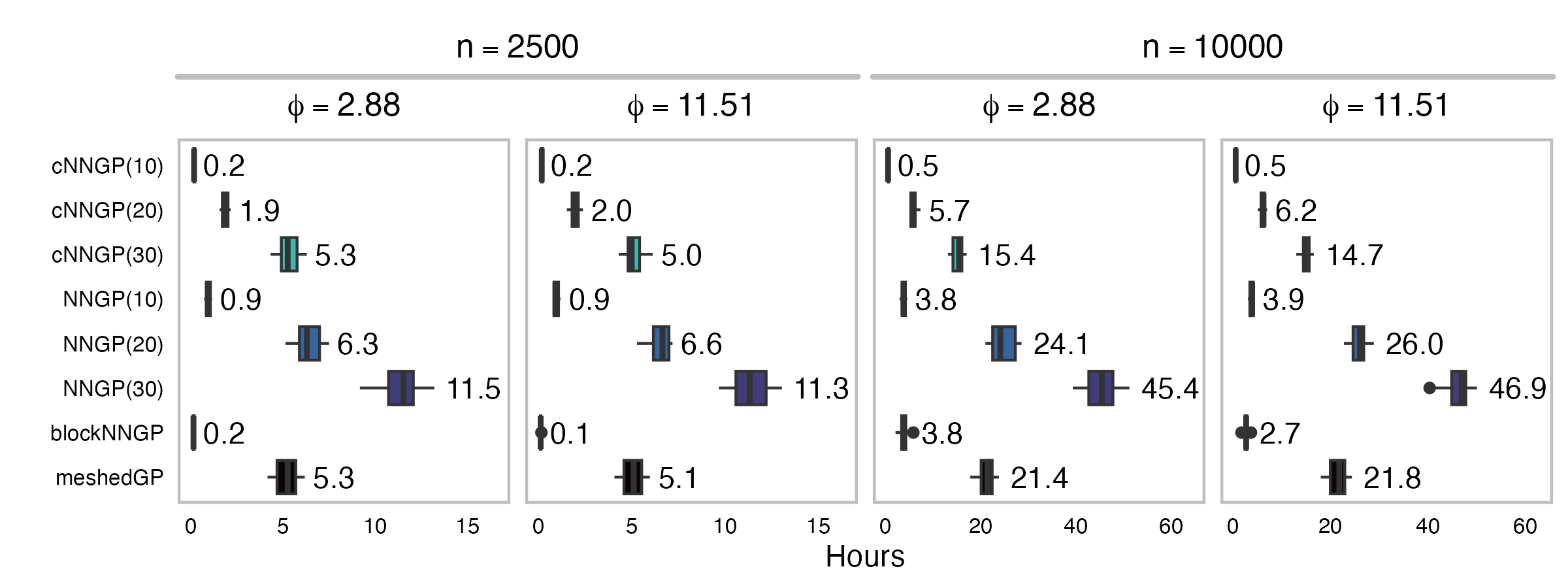}
    \caption{Fitting time in hours for each model. Boxplots show distribution of times for the 30 replicates of each ($n$,$\phi$) combination. The value beside each boxplot is the median fitting time in hours.}
    \label{fig:sim times}
\end{figure}

Model-fitting times are summarized in Figure \ref{fig:sim times}. For large datasets, the cNNGP with $m=10$ neighbors is the fastest method, outperforming the NNGP by reducing redundant covariance computations. For smaller datasets and the smaller neighbor set, it is competitive with the blockNNGP, which throughout all scenarios proved to be both efficient and inferentially on par with the NNGP. The computational advantage for the cNNGP grows with $n$, reflecting the increasing presence of repeated local distance patterns that the clustering algorithm exploits efficiently. This is remarkable considering that the blockNNGP uses INLA, whereas the cNNGP is MCMC based. 

\paragraph{Parameter Estimation}

\begin{figure}[ht]
 \centering
   \includegraphics[width=\textwidth]{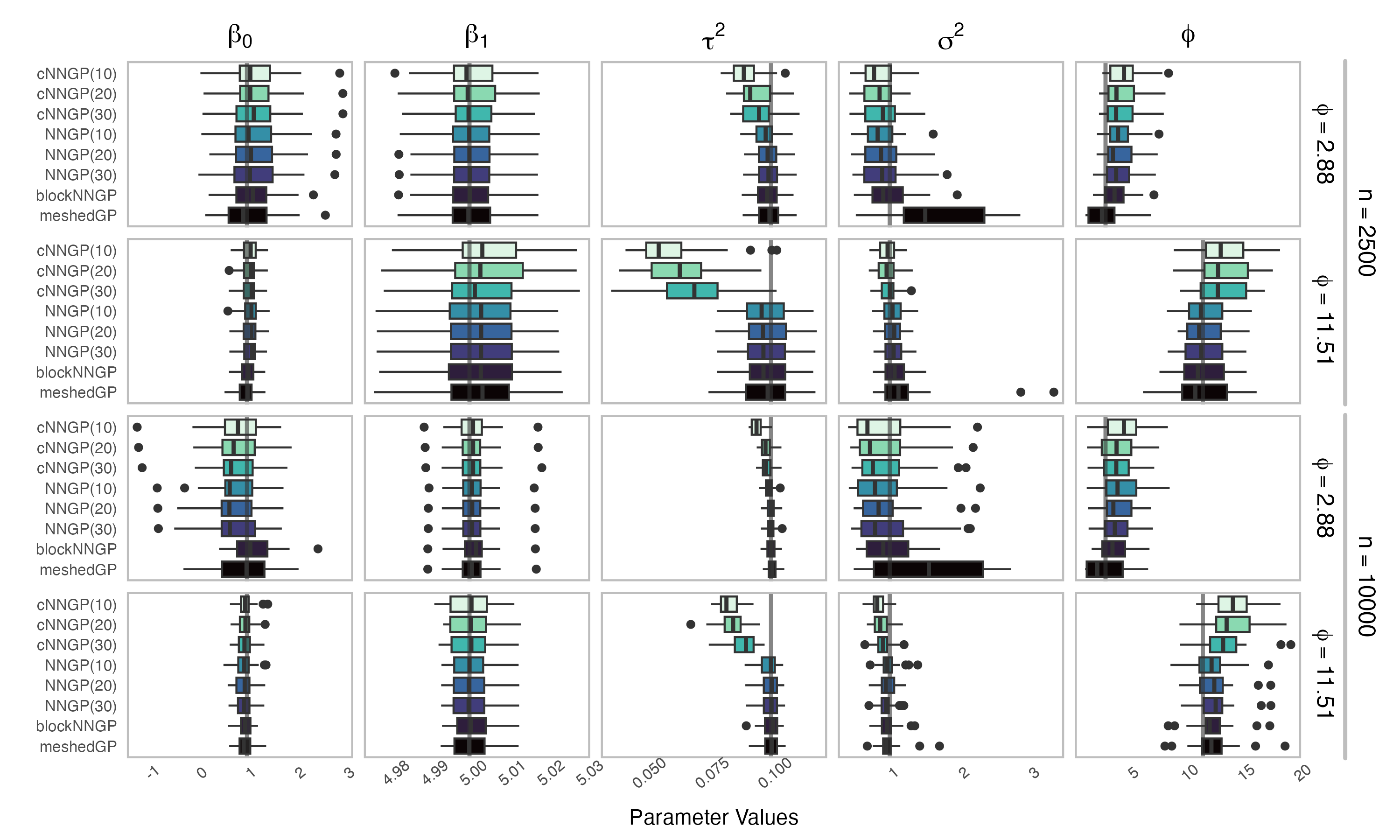}
   \caption{Box-plots for the parameter posterior means obtained from each model over the 30 replicates, for different sample sizes ($n\in\{2500, 10000\}$) and spatial ranges ($\phi\in\{2.88, 11.51\}$).}
   \label{fig:par_est}
\end{figure}

Figure \ref{fig:par_est} displays the distribution of parameter posterior means across the 30 replicated datasets for each $(n,\phi)$ combination. All models accurately recover the regression coefficients $\boldsymbol{\beta}$, with negligible bias. Differences arise primarily in the covariance parameters, and these discrepancies are most pronounced when the true spatial range is short.

Under short–range scenarios ($\phi$=11.51), where correlation decays quickly with distance, the covariance function is highly sensitive to discrepancies between a site’s true local distance vector $\bd_{\bs}$ and the cluster representative produced by the clustering algorithm. The cNNGP models exhibit the largest bias in $\tau^2$, typically underestimating it, and tends to overestimate $\phi$. Under the long–range scenario ($\phi=2.88$) the covariance changes slowly with distance, so modest deviations between $\bd_{\bs}$ and the cluster representative introduce relatively small error. Accordingly, the cNNGP models more closely match the NNGP and blockNNGP in its estimation of $\tau^2$, $\sigma^2$ and $\phi$. In the long spatial range scenarios, the MGP tends to overestimate $\sigma^2$ but it yields a more accurate point estimate for $\phi$. 

\begin{figure}[h]
 \centering
   \includegraphics[width=\textwidth]{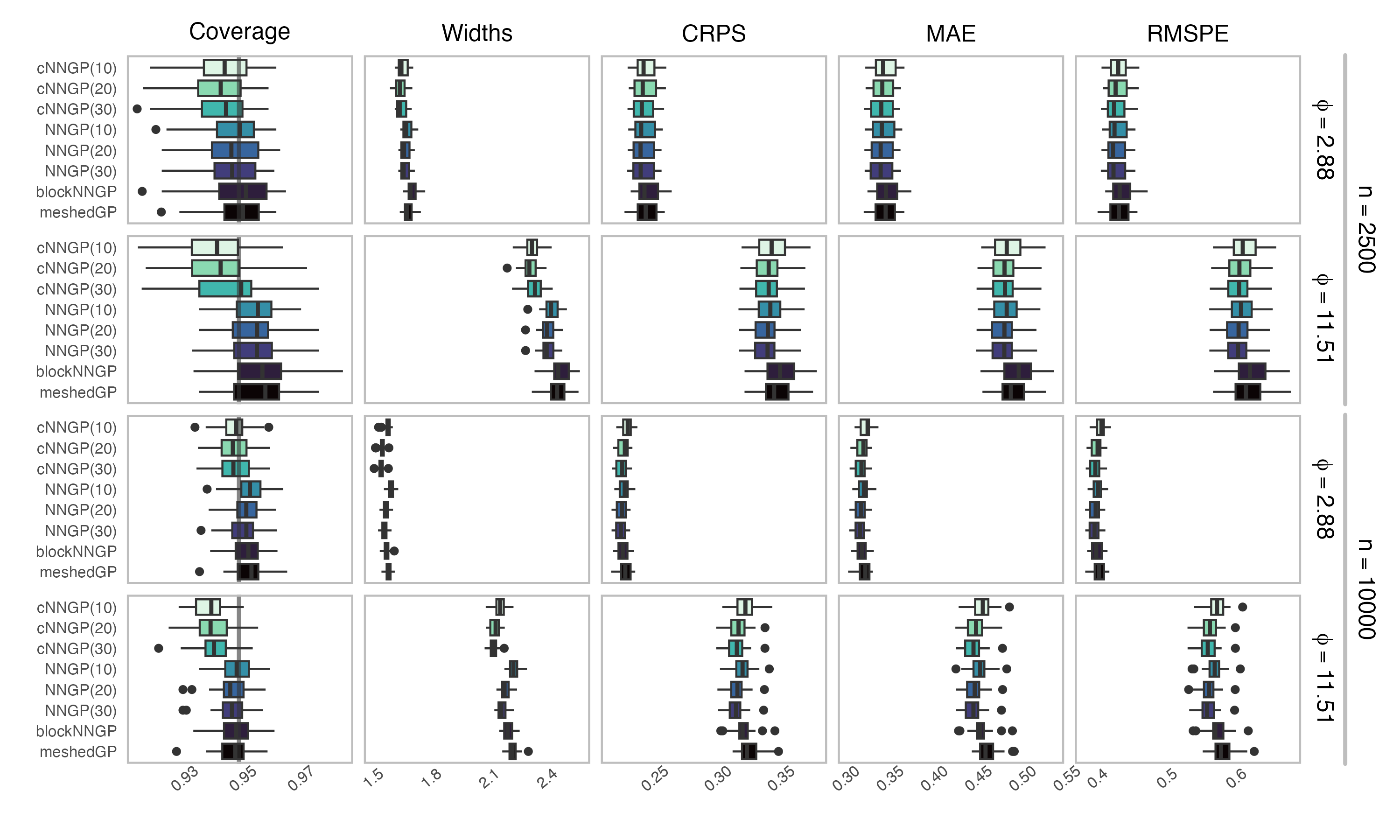}
   \caption{Box-plots for the prediction performance metrics considered for each model over the 30 replicates, for different sample sizes ($n\in\{2500, 10000\}$) and spatial ranges ($\phi\in\{2.88, 11.51\}$). These metrics include the coverage for out-of-sample predictions (Coverage), the median widths of the predictive intervals (Widths), the Continuous Ranked Probability Score (CRPS), the Mean Absolute Error (MAE) and the Root Mean Squared Prediction Error (RMSPE).}
   \label{fig:pred_perf}
\end{figure}

\paragraph{Uncertainty Quantification}

Figure \ref{fig:sim cov} shows credible-interval widths and empirical coverage for the model parameters. Estimation of $\beta_1$ is stable across all methods, with narrow intervals and coverage above 90\%. For $\beta_0$, blockNNGP and MGP occasionally produce wider intervals under long–range dependence. Differences are most pronounced for $\tau^2$ under the short–range scenario with the cNNGP variants, which is consistent with the amplification of distance-vector approximation error. For $\sigma^2$, the MGP exhibits the widest and most variable intervals.

\paragraph{Predictive Performance}

Despite the parameter differences, predictive accuracy is relatively similar across models (Figure \ref{fig:pred_perf}). Predictive coverage for the response $\by$ is close to the nominal 95\% confidence in all settings, with medians between 94\%–96\%. Continuous ranked probability score (CRPS), the mean absolute error (MAE) and the RMSE show minimal differences across models, with both the blockNNGP and the MGP doing slightly worse in the large sample size/small range scenario. The Watanabe–Akaike information criterion (WAIC) (Figure \ref{fig:sim waic}) confirms these patterns.

Crucially, even under the short–range scenario (where the cNNGP shows the largest parameter bias) its predictive performance is virtually indistinguishable from that of the NNGP. This reflects the fact that prediction depends primarily on accuracy in the local conditional distribution, which is preserved even when cluster representatives introduce moderate discrepancies in the global covariance structure.

\section{Application: Biomass Prediction using GEDI Data} \label{sec:gedi}

\begin{figure}
    \centering
    \subfigure{\includegraphics[width=0.48\textwidth]{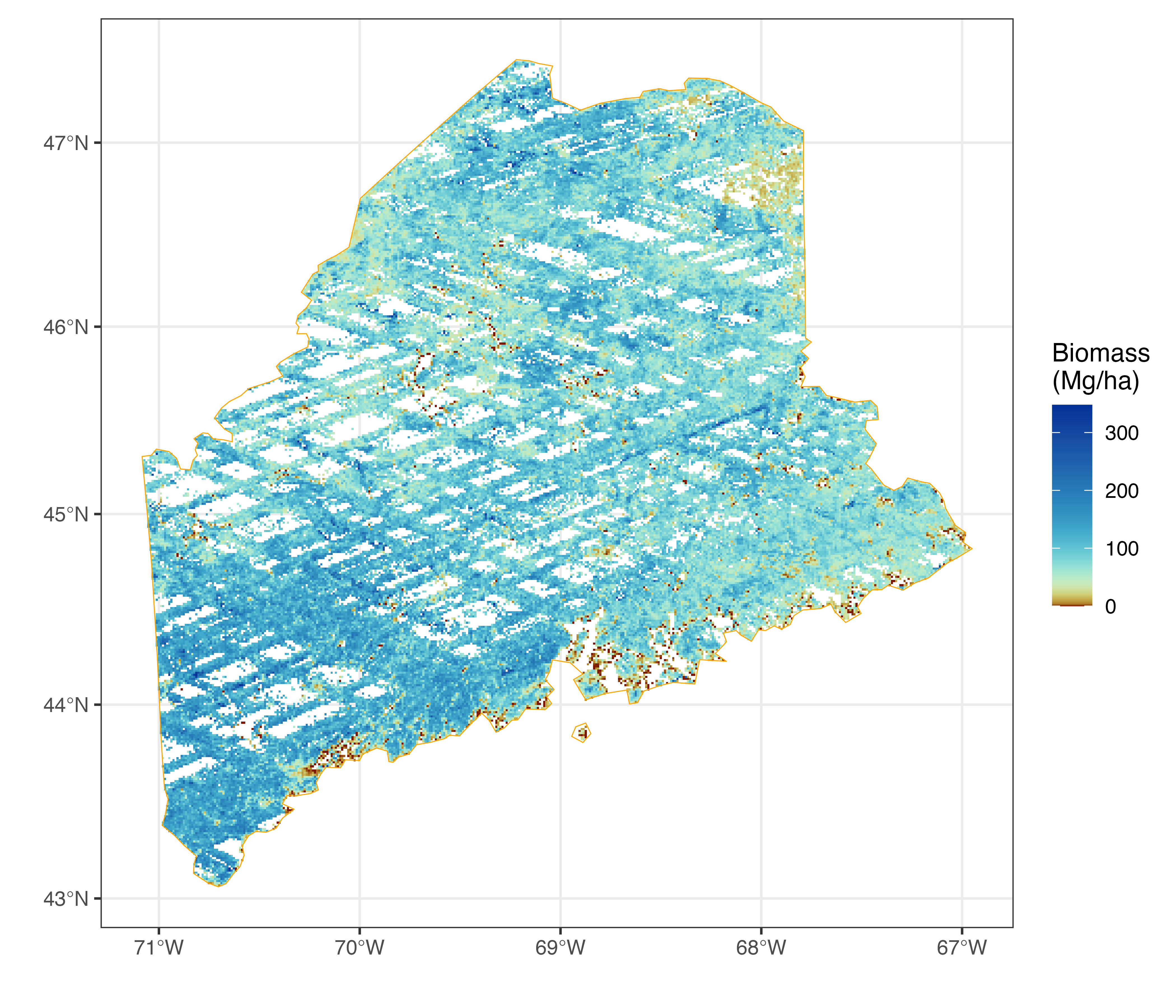}} 
    \subfigure{\includegraphics[width=0.48\textwidth]{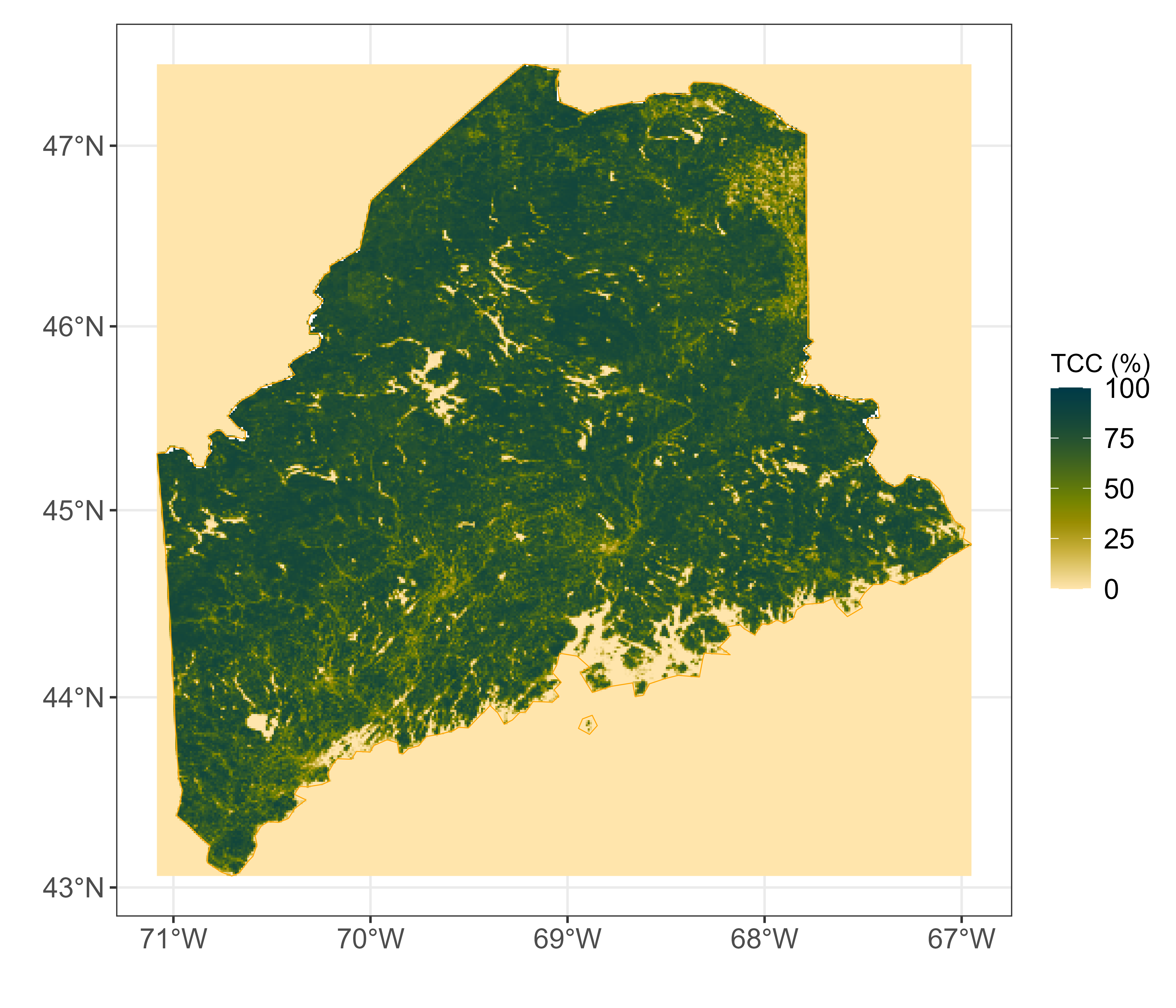}}
    \caption{GEDI data for state of Maine. \textbf{(Left):} Average above ground biomass from GEDI data. \textbf{(Right):} Proportion of tree canopy cover.}
    \label{fig:gedi}
\end{figure}

The Global Ecosystem Dynamics Investigation (GEDI) provides high-resolution lidar measurements of Earth’s forest structure to support global carbon-cycle science. The GEDI instrument is a geodetic-class laser ranging system installed on the International Space Station (ISS) in 2018, designed to sample approximately 4\% of Earth’s land surface between $51.6^\circ$ N and S latitude with a nominal 25-m footprint \citep{dubayah2020global}. For this analysis, we use the GEDI L4B data product, which provides global mean aboveground biomass density at a 1-km resolution between $52^\circ$ N and S, derived from observations collected between April 18, 2019 and March 16, 2023. Because GEDI samples along the ISS orbit track, data coverage becomes increasingly irregular at higher latitudes, leaving substantial gaps in regions such as Maine (Figure~\ref{fig:gedi}). Our objective is to use the cNNGP to spatially predict biomass in these unsampled areas and produce wall-to-wall estimates across the state, and contrast these results with those obtained using the NNGP. In addition to leveraging spatial dependence among observed GEDI pixels, we incorporate a complete-coverage tree canopy cover (TCC) dataset as a covariate \citep{tcc_methods_2023}. Figure~\ref{fig:gedi} displays the datasets used in this analysis as well as Figures \ref{fig:bio-tcc-dist} and  \ref{fig:bio-vs-tcc} in the appendix.

At the 1 km by 1 km resolution, there are 86,949 pixels in Maine. We removed any values of TCC over 100\%, leaving 86,802 pixels. Of these, 16,914 pixels did not have recorded biomass values, and we deleted 112 biomass values that were unrealistically high (above 350 Mg/ha) for Maine. The coordinates were scaled accordingly to have distances be measured in kilometers. Pixel centroids were used to compute distances in subsequent models. The grid of hexagons was 30 by 30 over the state of Maine. As in our simulations, we used the \textsf{blockCV} package to divide locations into 5 folds, one of which was held out for comparing the methods' predictive performance, see Figure \ref{fig:gedi holdouts}. In summary, a total of 56,996 pixels used for model training, the hold-out set used for model validation consisted of 12,780 pixels with known biomass. Prediction was also done at 17,026 additional pixels that had no biomass measurement.

For both the cNNGP and the NNGP we set the number of neighbors to $m=20$. To identify a suitable clustering radius for the cNNGP we first apply PCA to the matrix of neighbor set distances given by $(\bd_{\bs_1},\ldots, \bd_{\bs_n})'$, reducing the number from ${m+1\choose 2}=210$ to 35 columns that explained 90\% of the variability of the original data. From this reduced matrix we sampled at random 10,000 rows (i.e., locations). Then we apply the leader clustering procedure to the subsample for radius values $r=1,2,\ldots,10$. Using the resulting ($r$, $\kappa$) pairs, we identify the elbow in the $\kappa$ vs. $r$ curve and the selected radius value is used for all locations. Using PCA to reduce the dimensionality of the matrix and then subsampling a manageable set of observations greatly reduces the computational cost of tuning. 

This preliminary analysis indicated that a radius $r=4$ offered a good balance between within-cluster homogeneity and the total number of clusters (see Figure \ref{fig:gedi kappa} in the supplement). We then applied the leader algorithm with $r=4$ to the full dataset, yielding 3,671 clusters, each associated with a unique representative distance matrix—only 6.4\% of the matrices needed under the full NNGP. These 3,671 sets of distance-based NNGP factors were subsequently used for model fitting.

For choosing the prior for $\phi$, 10,000 locations were subsampled and the minimum $d_{min}$ and maximum $d_{max}$ distances in this subset of locations were used to assign a uniform prior with lower bound $3/d_{max}$ and upper bound $3/d_{min}$, resulting in a $\text{Uniform}(0.01,3)$ prior. Flat priors were set for the intercept and TCC regression coefficients. The priors for $\sigma^2$ and $\tau^2$ were both chosen to be $\text{InvGamma}(2,600)$, where the hyperparameter values were selected to conform to the scale of the response.

Each model was fit using three chains of 30,000 samples. The chains were initialized with different random seeds and making use of the same approach to chose starting values as the one used with the simulated data. All three chains indicated good convergence and stabilized rapidly, which is why only the first 500 samples of each chain were discarded as burn-in (see Figures \ref{fig:gedi_traceplot_theta}, and \ref{fig:gedi_traceplot_w}, \ref{fig:gedi_traceplot_beta} for traceplots).  Both models (NNGP and cNNGP) were fit with 5 \textsf{OpenMP} threads on the same machine used for our simulation study. 

\begin{table}[t]
    \centering
    \begin{tabular}{|r|rc|rc|}
\hline
{} & \multicolumn{2}{c|}{cNNGP} & \multicolumn{2}{c|}{NNGP} \\ 
\cline{2-3} \cline{4-5}
Parameter & Mean & Interval & Mean & Interval \\
\hline
Intercept & 26.26 & (24.09, 28.51) & 26.31 & (24.11, 28.61) \\ 
  TCC & 0.97 & (0.95, 0.99) & 0.96 & (0.94, 0.98) \\ 
  $\sigma^2$ & 881.27 & (842.73, 920.18) & 920.94 & (878.14, 965.10) \\ 
  $\tau^2$ & 348.09 & (333.70, 362.78) & 373.42 & (359.80, 386.95) \\ 
  $\phi$ & 0.26 & (0.24, 0.28) & 0.26 & (0.24, 0.28) \\ 
   \hline
\end{tabular}
    \quad\quad 
    \begin{tabular}{|rrr|}
\hline
Metric & cNNGP & NNGP \\ 
  \hline
CRPS & 16.07 & 16.13 \\ 
  MAE & 22.31 & 22.34 \\ 
  RMSPE & 29.52 & 29.57 \\ 
  Coverage & 95.9\% & 96.3\% \\ 
  Widths & 123.15 & 126.52 \\ 
  Time (hours) & 177.34 & 896.46 \\
   \hline
\end{tabular}
    \caption{Posterior parameter means and credible intervals (left table), and out-of-sample prediction metrics (right table) for the GEDI data from the cNNGP and NNGP.}
    \label{tab:gedi params}
\end{table}

This analysis illustrates the practical advantages of the clustered NNGP and reinforces the insights gained from the simulation study. As shown in Table \ref{tab:gedi params}, the cNNGP yields parameter estimates that align extremely closely with the NNGP. The regression coefficients and the spatial decay parameter $\phi$ are nearly identical, demonstrating that the clustering procedure preserves the large-scale dependency structure captured by the NNGP. The only meaningful deviation occurs in the process variance $\sigma^2$, where the cNNGP estimate is approximately 4.5\% lower. This mirrors what we observed in the simulations: when locations share representative NNGP factors within clusters, very fine-scale variability is slightly smoothed, leading to modest underestimation of $\sigma^2$.

Importantly, for the exponential covariance considered, the estimated value of $\phi\approx 0.26$ implies a long spatial range. This setting is especially favorable for the cNNGP. Under long-range dependence, each conditional density is not as sensitive to smaller deviations between the true local distance matrices and the cluster-representative matrix used for the approximation. As such, the cluster-based approximations effectively capture these recurring long-range patterns across locations with a relatively small approximation error. This is precisely the regime where our simulation study showed that the cNNGP performs most similarly to the NNGP.

\begin{figure}[ht]
    \centering
    \includegraphics[width=0.8\linewidth]{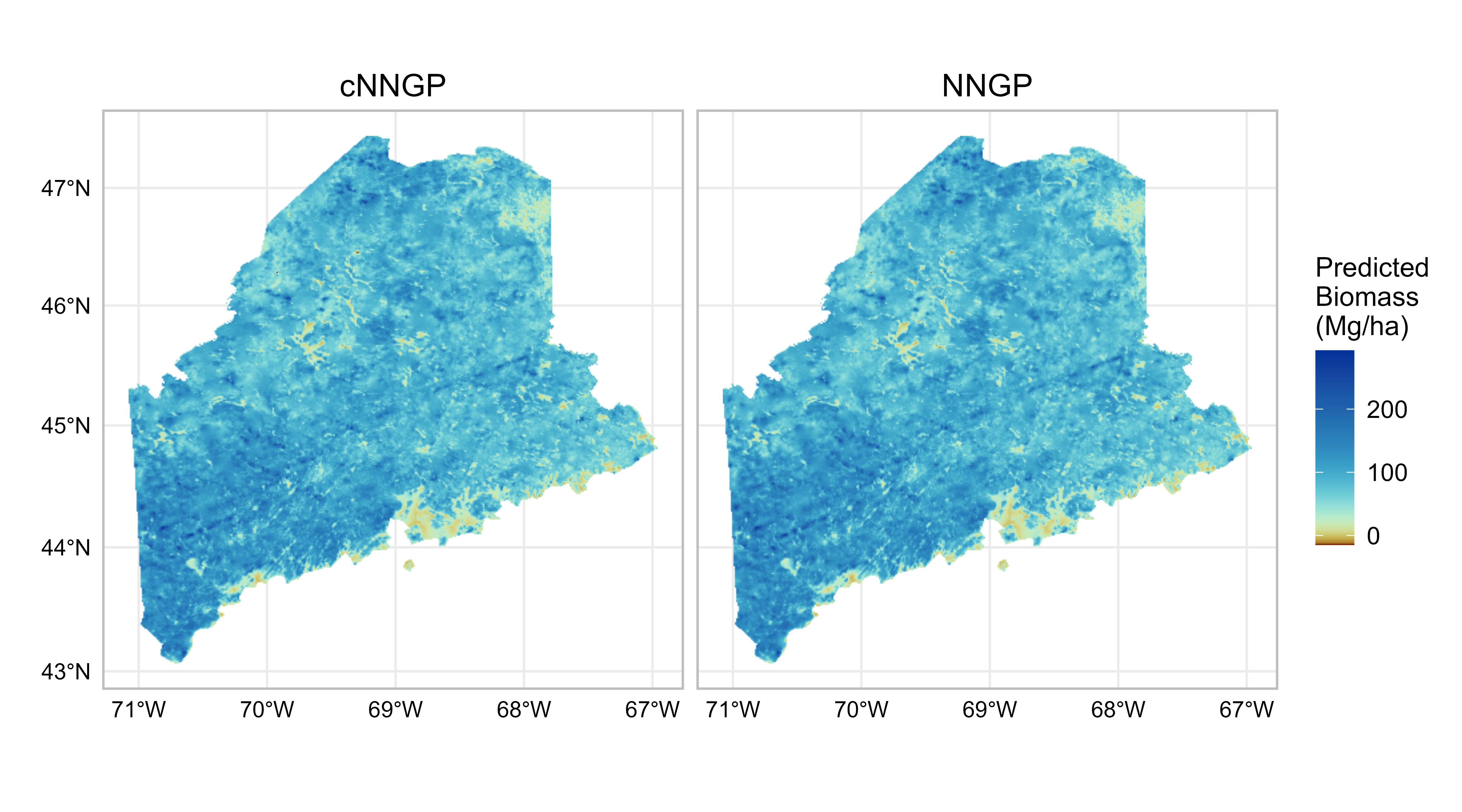}
    \caption{Predicted biomass values for all GEDI locations.}
    \label{fig:gedi means}
\end{figure}

The predictive results confirm this behavior. The predictions obtained by either method appear identical (Figure \ref{fig:gedi holdout prediction}). The posterior mean biomass maps and associated standard errors (Figures \ref{fig:gedi means} and \ref{fig:gedi se}) are visually indistinguishable across models, reflecting the cNNGP’s ability to retain the essence of the dependence structure. The differences between the true and predicted biomass values are included in Figure \ref{fig:gedi est-true} where the cNNGP and NNGP again provide visually equivalent maps. The differences between the cNNGP and NNGP posterior means and standard errors are seen in Figures \ref{fig:gedi mean diff} and \ref{fig:gedi se diff} which show that the cNNGP tends to have wider credible intervals on the fitting data but more certainty in predictions. Holdout prediction coverage is also nearly identical—95.9\% for the cNNGP versus 96.3\% for the NNGP—and the median credible interval widths differ by only 3.4 units (123.15 vs. 126.52). Point-prediction metrics (CRPS, MAE, RMSPE) match to about one decimal place, confirming that the slight differences in covariance parameters do not translate into meaningful losses in predictive accuracy.

The results underscore the central advantage of the cNNGP; dramatic computational savings with negligible loss in predictive fidelity. For this dataset, the clustering step reduced the number of unique neighbor-distance matrices to only 6.4\% of what is needed for the NNGP, lowering the computational cost from $\mathcal{O}(nm^3)$ to $\mathcal{O}(\kappa m^3)$ and cutting runtime from 896 hours to 177 hours (i.e., only 19.8\% of the time taken by the NNGP). The long spatial range of the GEDI signal amplifies these gains, since it provides precisely the structure that the cNNGP is best equipped to exploit.

\section{Discussion} \label{sec:discussion}

The NNGP has established itself as a powerful approximation to dense Gaussian processes and has enabled Bayesian spatial modeling for datasets that were previously computationally infeasible. However, the $\mathcal{O}(nm^3)$ complexity of the NNGP still poses challenges as the scale and resolution of modern spatial datasets continue to grow. The clustered NNGP (cNNGP) is an intuitive solution that directly addresses this bottleneck by exploiting recurring local spatial patterns: using a clustering step, locations with similar neighbor-distance structures are grouped, and a single cluster representative is used to construct the NNGP factors for all members of the cluster. This approach reduces redundant computations while preserving the essential local dependency structure needed for accurate prediction.

Across simulated datasets with 2,500 and 10,000 locations, despite its simplicity, the cNNGP delivered substantial computational gains while maintaining predictive performance essentially indistinguishable from the NNGP. In fact, for the larger datasets, the cNNGP with $m=10$ neighbors was the fastest method tested, outperforming even the blockNNGP, which is based on INLA and optimized for speed. The main source of approximation error in the cNNGP appears under short-range dependence, where the cNNGP tends to overestimate $\phi$ and underestimate $\tau^2$, leading to higher WAIC values relative to the NNGP. These discrepancies diminish under long-range dependence and, importantly, do not translate in either case into meaningful losses in predictive accuracy, even in the most challenging short-range scenarios. 

The real-data analysis using GEDI-derived biomass similarly underscores the practical utility of the cNNGP. Parameter estimates from the cNNGP closely matched those from the NNGP, holdout predictive coverage differed by less than 1\%, and credible intervals were slightly narrower. Crucially, the cNNGP achieved these results while requiring only one-fifth of the computational time of the NNGP. These findings highlight a central advantage of the cNNGP: when the number of clusters $\kappa$ produced by the leader algorithm is small relative to the number of locations $n$, the computational cost is reduced from $\mathcal{O}(nm^3)$ to $\mathcal{O}(\kappa m^3)$, yielding substantial speedups with only modest loss of inferential precision. The value of $\kappa$ depends on the heterogeneity of local neighbor-distance patterns and the choice of clustering radius $r$, which allows users to balance computational efficiency and approximation fidelity.

At the same time, the cNNGP inherits the assumptions of the NNGP and imposes new constraints due to the reliance on the similarities among distance sets of nearest neighbor locations for clustering. Because of these constraints, the method currently applies only to stationary and isotropic processes; extending it in these directions would require alternative definitions of similarity among local covariance patterns. Further, being an approximation to the NNGP, the cNNGP is appropriate only in settings where the NNGP itself is a suitable modeling framework.

Overall, the cNNGP offers a straightforward alternative for practitioners seeking fast, and scalable spatial inference with minimal degradation in predictive performance. By leveraging repeating local structure in large spatial datasets, it enables substantial computational savings while remaining faithful to the core predictive strengths of the NNGP. Future work may expand the approach to nonstationary settings, alternative clustering strategies, or adaptive selection of radius parameters to automate the tradeoff between speed and accuracy.

\section*{Acknowledgements}

The authors would like thank Michele Peruzzi for his generous input defining the settings for the MGP used in our simulation experiments. This work was partially supported by the U.S. Department of Energy, Office of Science, Office of Advanced Scientific Computing Research Award DOE CSGF DE-SC0023112, the United States Forest Service, the National Science Foundation Awards RTG DMS-2136228 and DEB-2213565, and the National Aeronautics and Space Administration Carbon Monitoring system grants Hayes 2020 and 2023.

\section*{Disclaimer}

This report was prepared as an account of work sponsored by an agency of the
United States Government. Neither the United States Government nor any agency thereof, nor
any of their employees, makes any warranty, express or implied, or assumes any legal liability
or responsibility for the accuracy, completeness, or usefulness of any information, apparatus,
product, or process disclosed, or represents that its use would not infringe privately owned
rights. Reference herein to any specific commercial product, process, or service by trade name,
trademark, manufacturer, or otherwise does not necessarily constitute or imply its
endorsement, recommendation, or favoring by the United States Government or any agency
thereof. The views and opinions of authors expressed herein do not necessarily state or reflect
those of the United States Government or any agency thereof.

\bibliographystyle{unsrtnat}
\bibliography{references}

\clearpage

\section{Appendix}
\renewcommand{\thefigure}{A\arabic{figure}}
\setcounter{figure}{0}

The appendix includes two additional figures from the analysis of the simulated datasets and twelve figures for the GEDI data analysis.

\subsection{Additional figures from simulations}

\begin{figure}[h]
    \centering
    \includegraphics[width=0.95\linewidth]{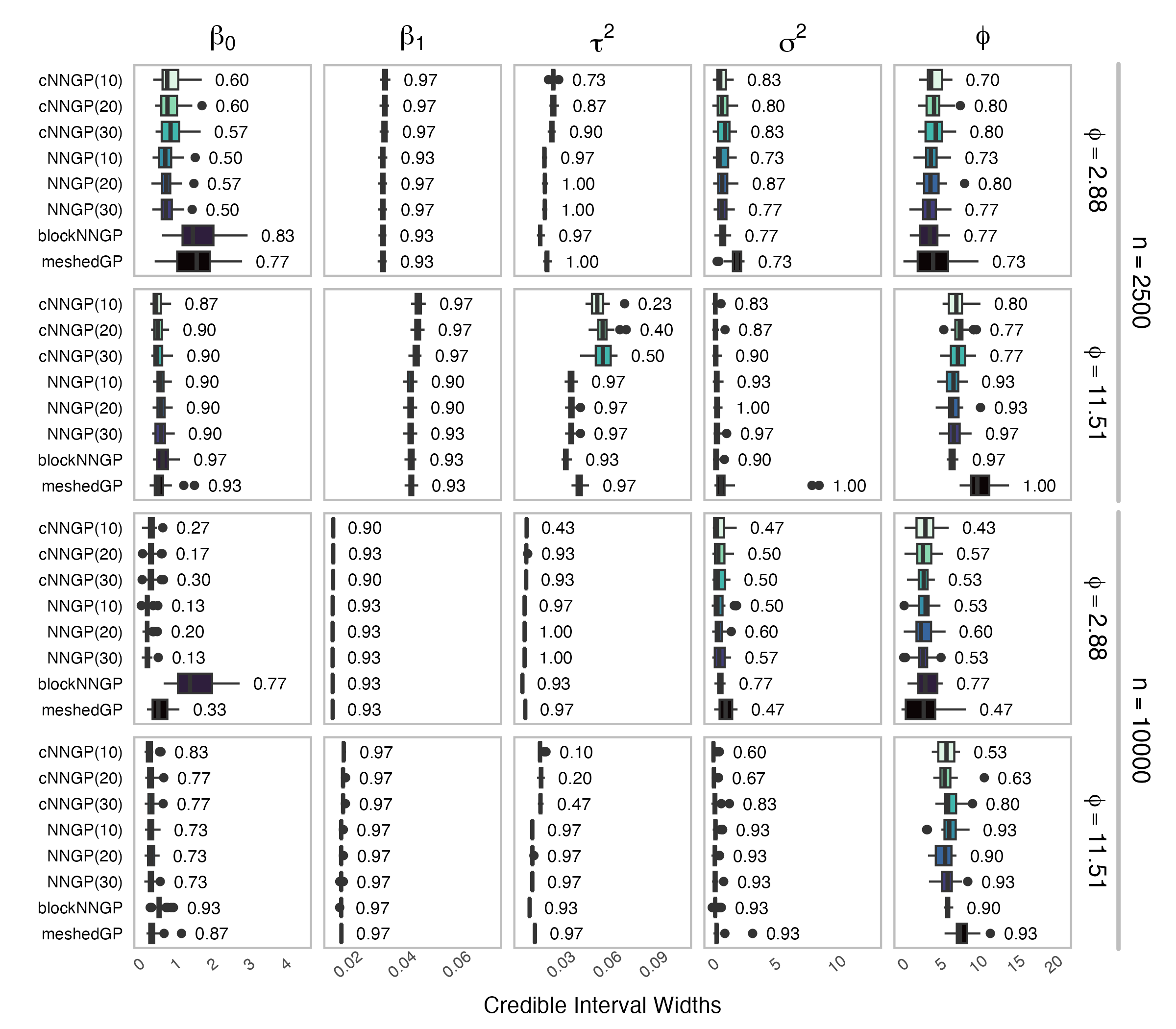}
    \caption{$\,$ Boxplots for the credible interval widths for all model parameters under each model over the 30 dataset replicates. The number shown beside each boxplot is the proportion of intervals which captured the true parameter value over the 30 replicates.}
    \label{fig:sim cov}
\end{figure}

\begin{figure}
    \centering
    \includegraphics[width=0.8\linewidth]{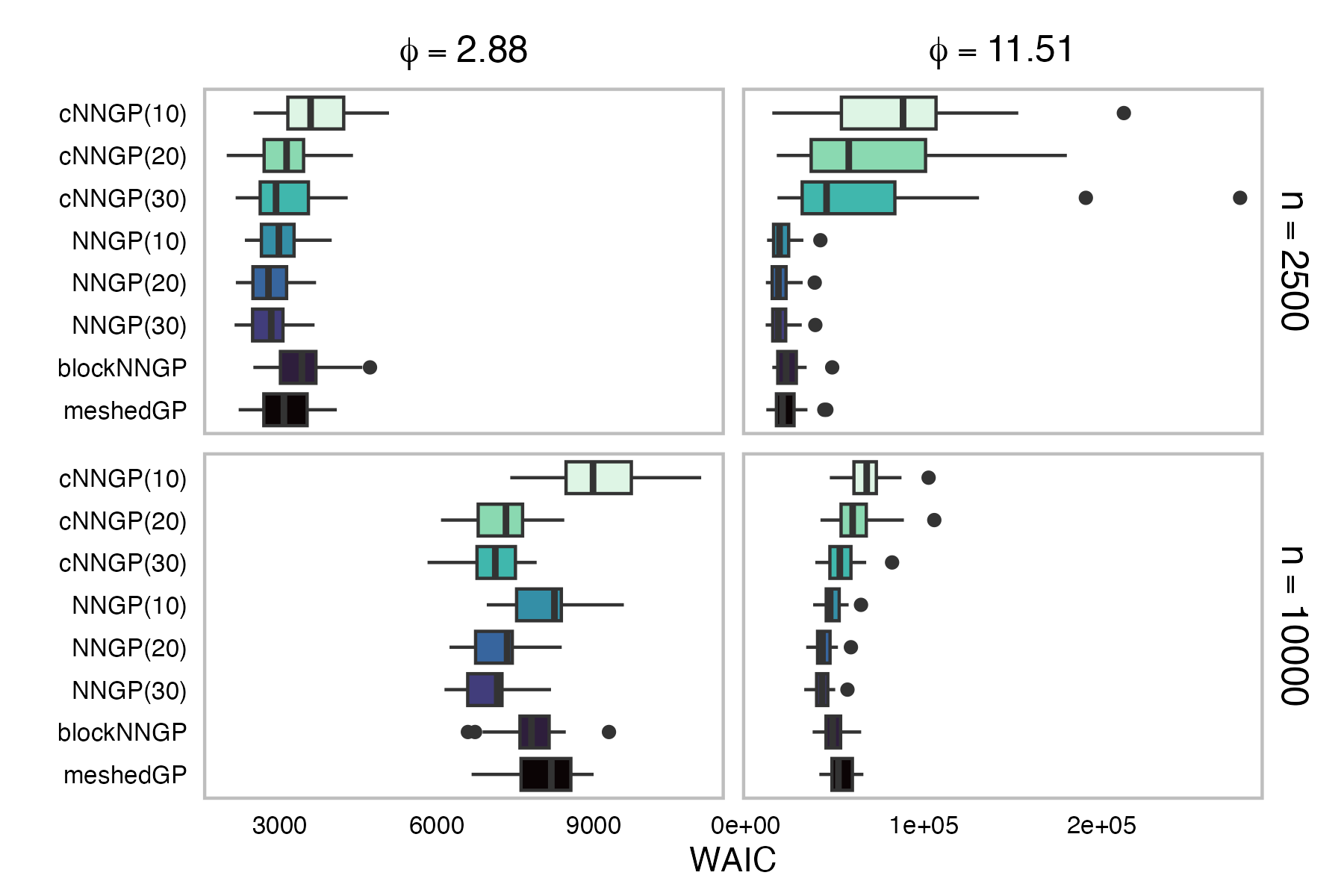}
    \caption{$\,$ Boxplots for the WAIC for each model over the 30 dataset replicates.}
    \label{fig:sim waic}
\end{figure}

\clearpage

\subsection{Additional figures GEDI data analysis}

\begin{figure}[h]
    \centering
    \includegraphics[width=0.8\textwidth]{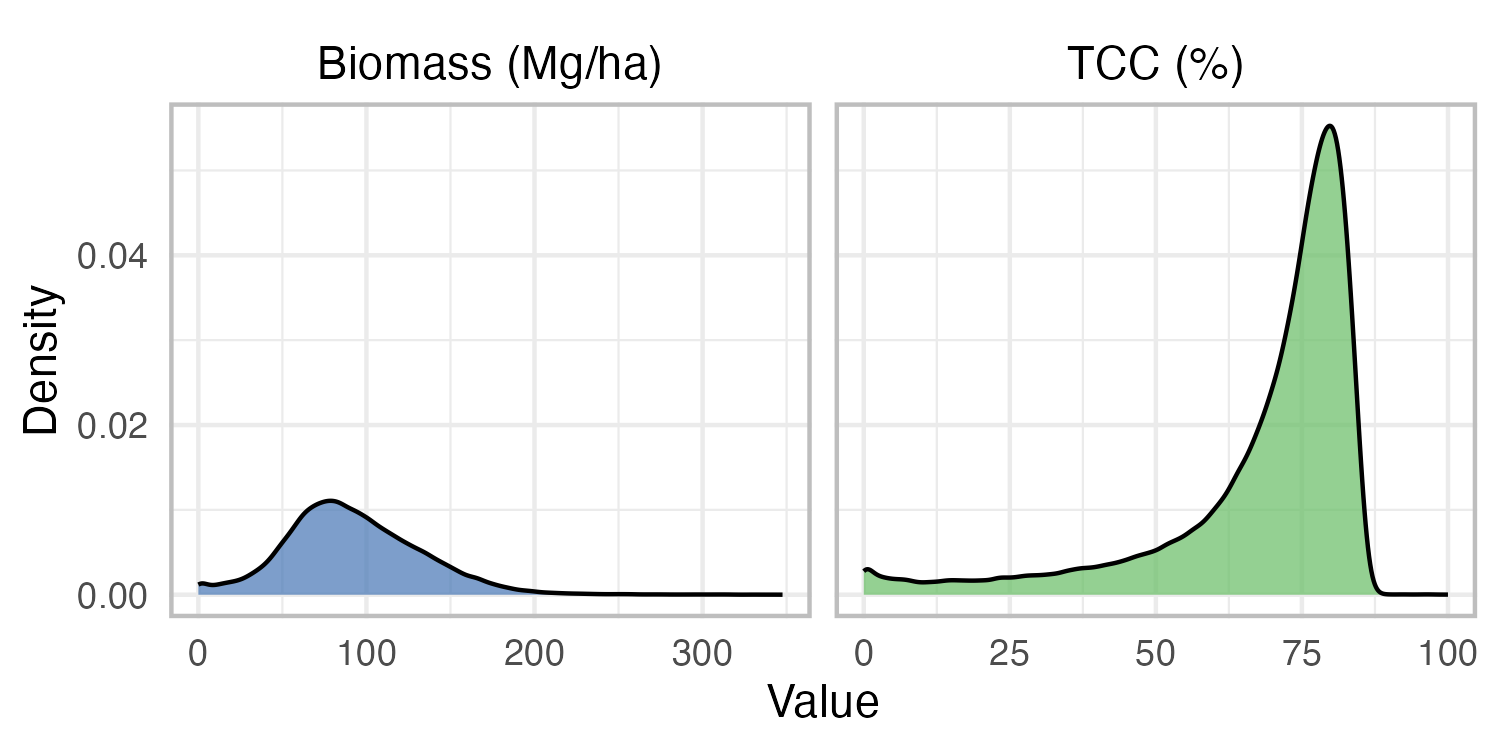}
    \caption{$\,$ Distribution of biomass and total canopy cover over the state of Maine.}
    \label{fig:bio-tcc-dist}
\end{figure}

\begin{figure}
    \centering
    \includegraphics[width=0.7\textwidth]{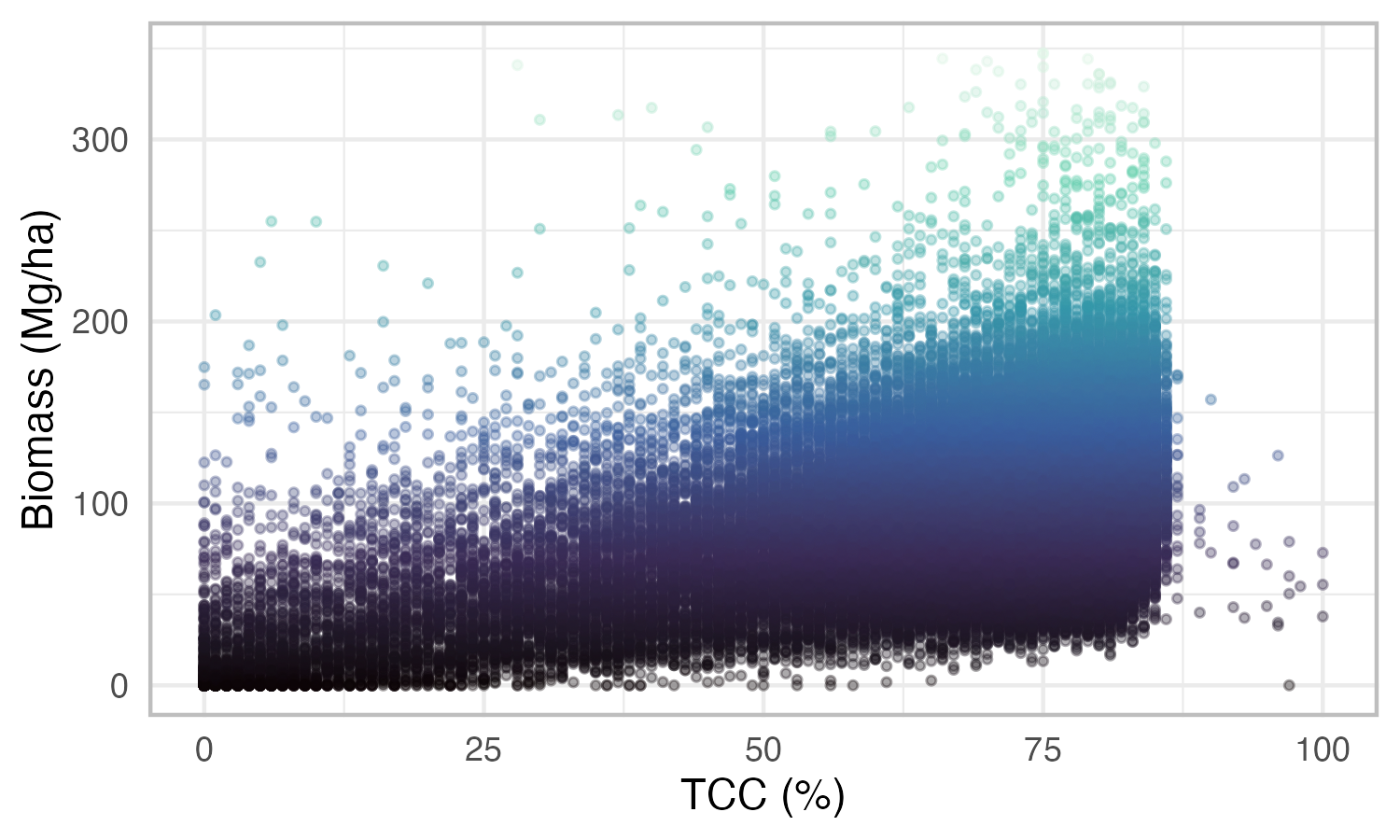}
    \caption{$\,$ Scatterplot of biomass against total canopy cover over the state of Maine.}
    \label{fig:bio-vs-tcc}
\end{figure}

\begin{figure}
    \centering
    \includegraphics[width=0.5\textwidth]{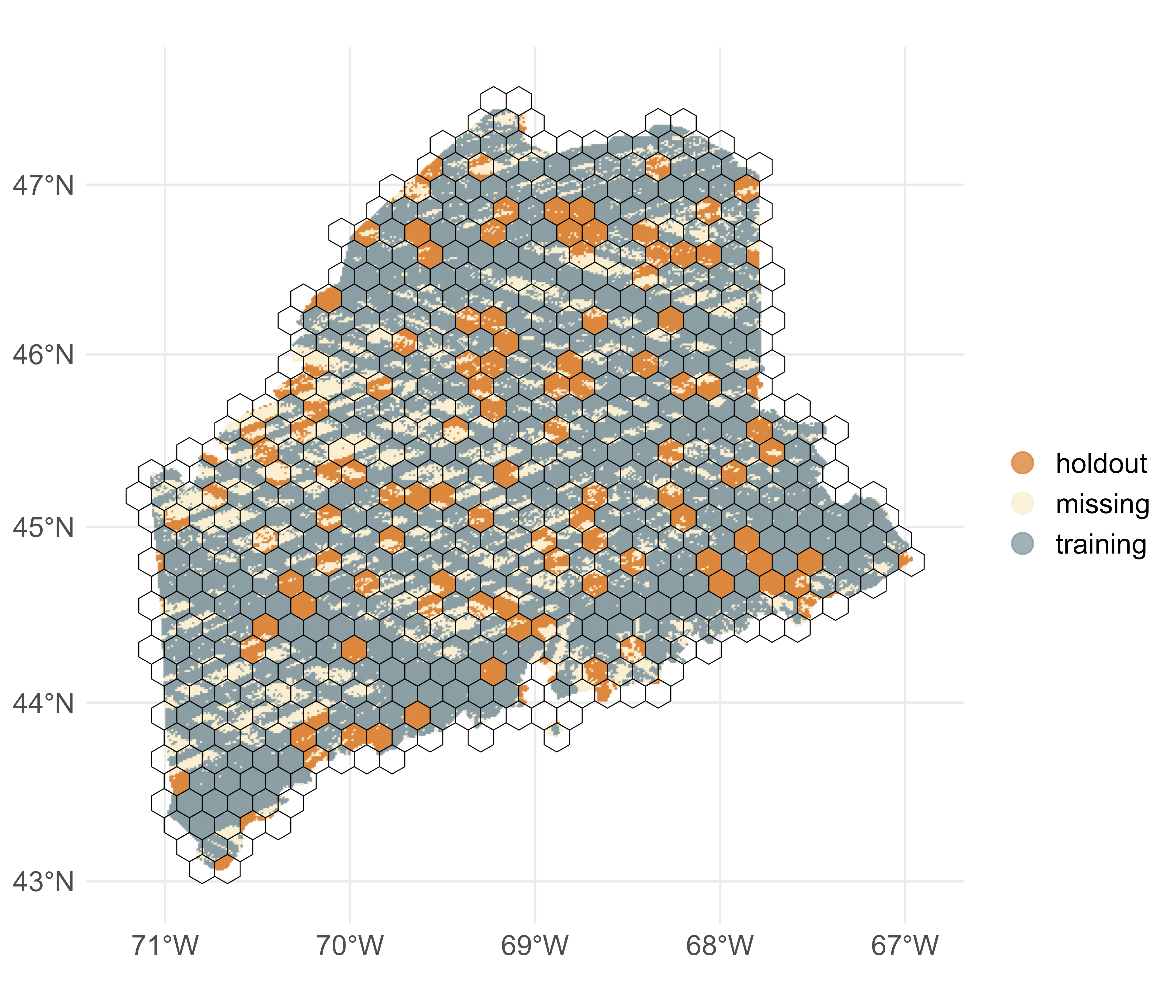}
    \caption{$\,$ Assignment of GEDI locations used for model training, locations where the true biomass value is known but held out to assess prediction performance, and locations where the true biomass value is unknown (shown as 'missing' in the legend).}
    \label{fig:gedi holdouts}
\end{figure}

\begin{figure}[h]
    \centering
    \includegraphics[width=0.5\textwidth]{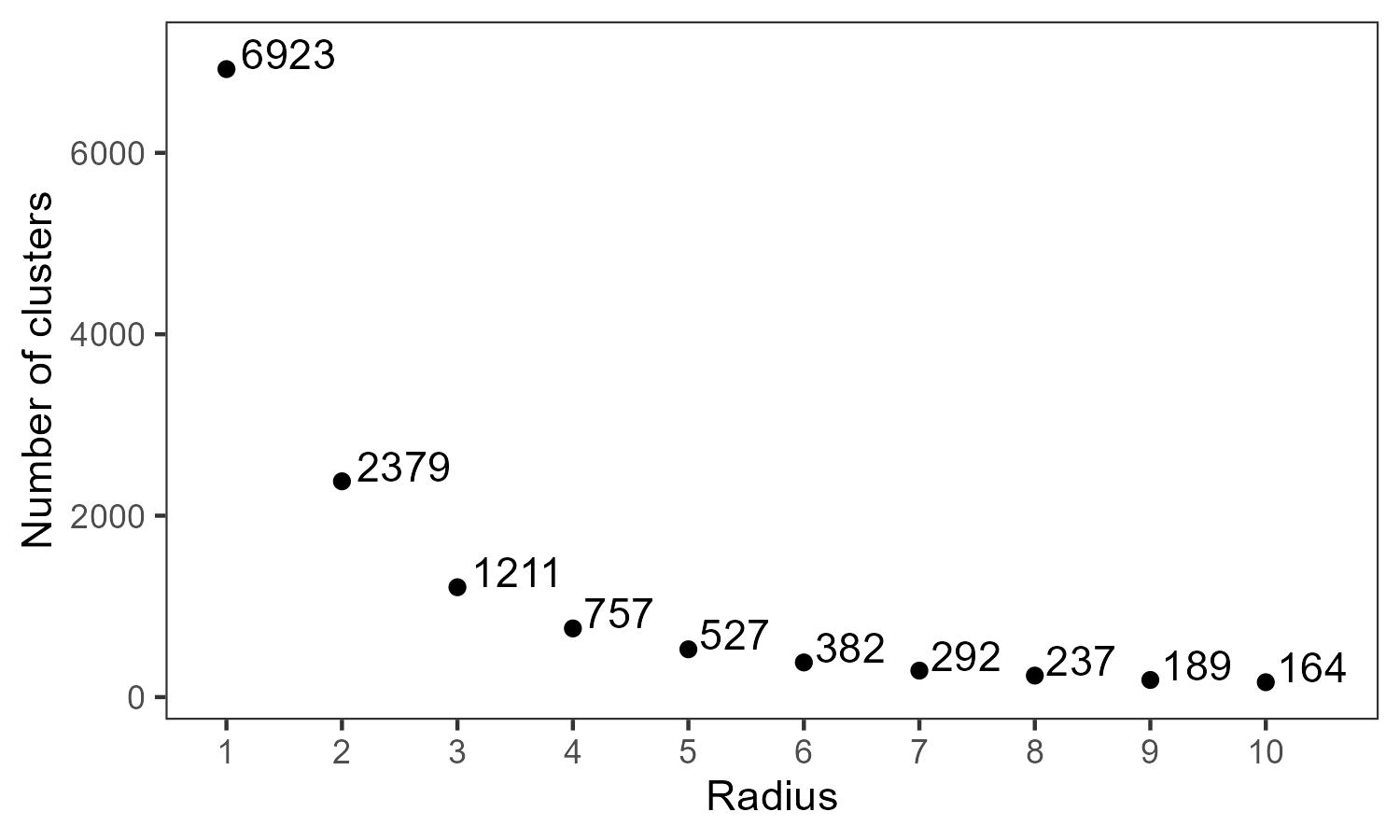} 
    \caption{$\,$ Number of clusters vs. clustering radius for subsample of size 10,000 from the GEDI data. The radius value of 4 was selected for the analysis.}
    \label{fig:gedi kappa}
\end{figure}

\begin{figure}
    \centering
    \includegraphics[width=0.8\textwidth]{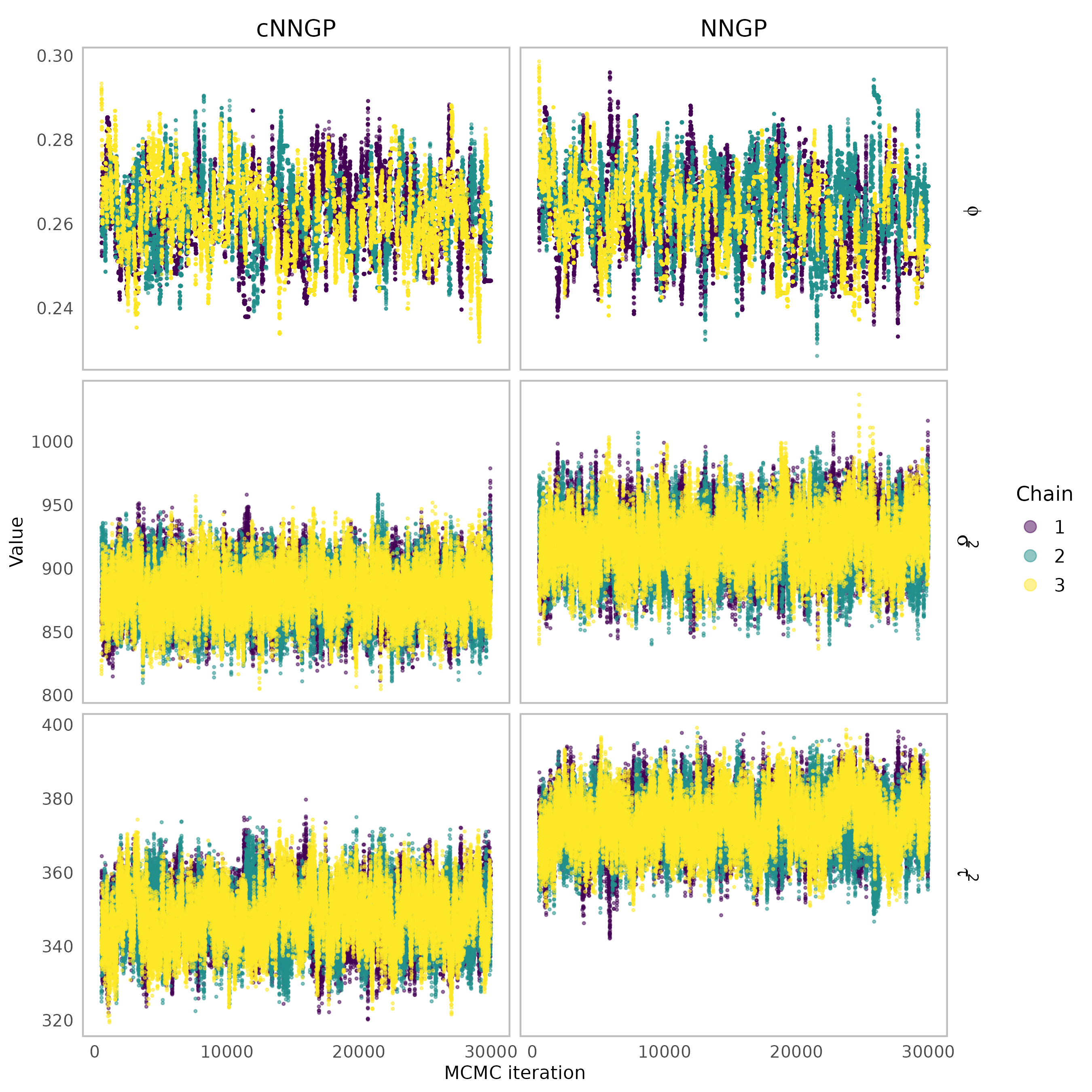} 
    \caption{$\,$ GEDI traceplot of covariance parameters from the cNNGP model. The first 500 samples (burn-in period) are omitted for figure clarity.}
    \label{fig:gedi_traceplot_theta}
\end{figure}

\begin{figure}
    \centering
    \includegraphics[width=0.8\textwidth]{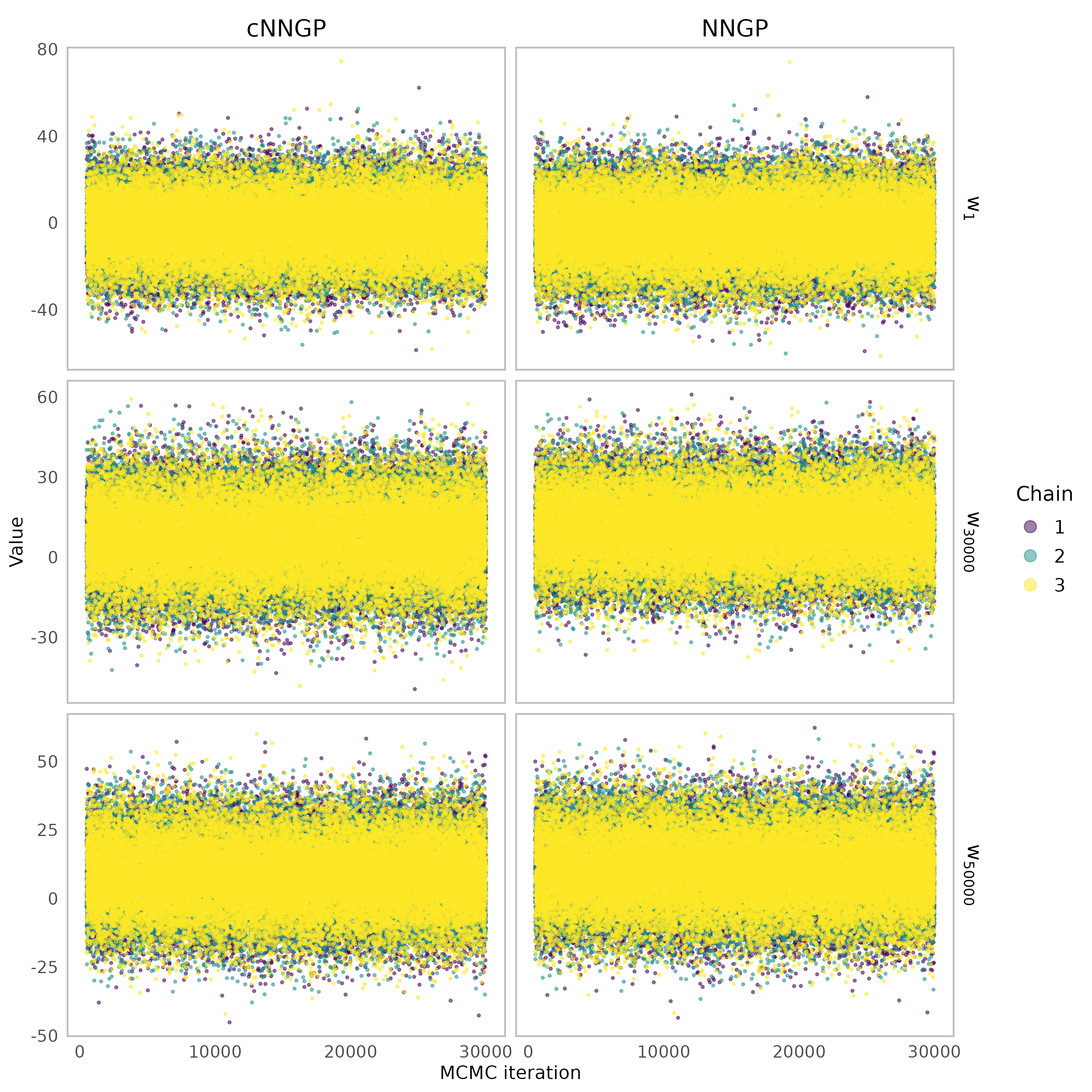} 
    \caption{$\,$ GEDI traceplot from the cNNGP model for the spatial effects $w_1$, $w_{30000}$ and $w_{50000}$ corresponding to the ordered locations $\bs_{1}, \bs_{30000}, \bs_{50000}$. The first 500 samples (burn-in period) are omitted for figure clarity.}
    \label{fig:gedi_traceplot_w}
\end{figure}

\begin{figure}
    \centering
    \includegraphics[width=0.8\textwidth]{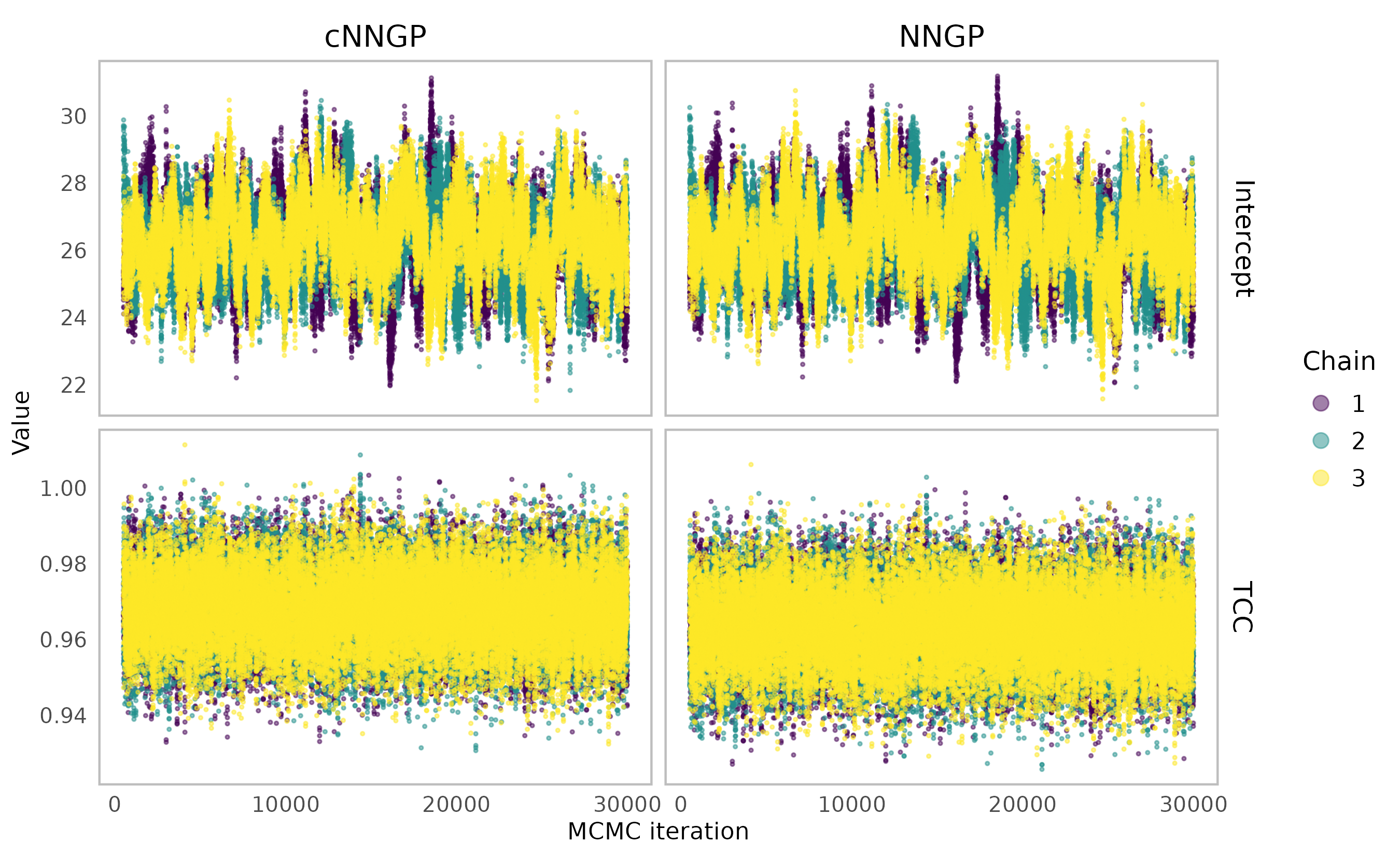} 
    \caption{$\,$ GEDI traceplot of the regression coefficients from the NNGP and the cNNGP models. The first 500 samples (burn-in period) are omitted for figure clarity.}
    \label{fig:gedi_traceplot_beta}
\end{figure}

\begin{figure}
    \centering
    \includegraphics[width=0.9\textwidth]{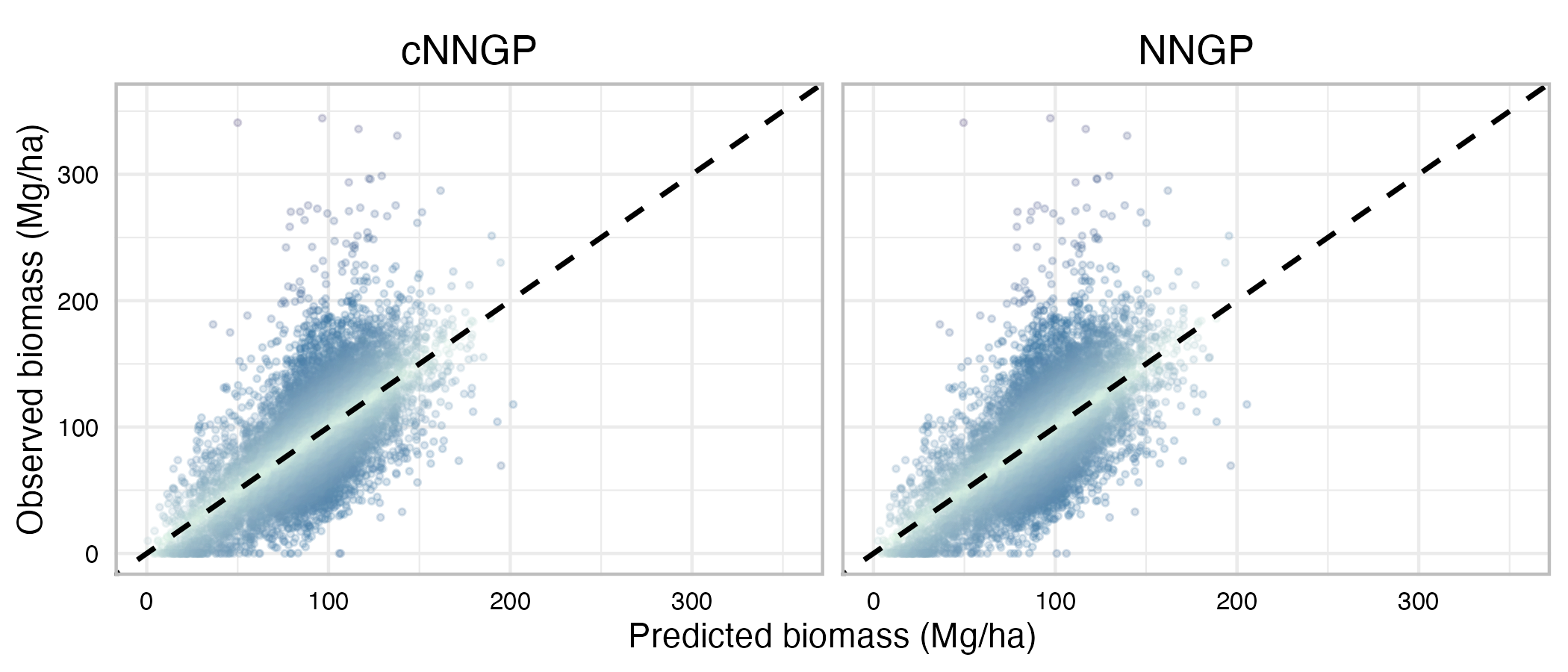} 
    \caption{$\,$ Observed versus predicted biomass from holdout data for each model. The correlation between the posterior predictive mean and true biomass is 65.7\% for the cNNGP and 65.5\% for the NNGP.}
    \label{fig:gedi holdout prediction}
\end{figure}

\begin{figure}
    \centering
    \includegraphics[width=0.9\linewidth]{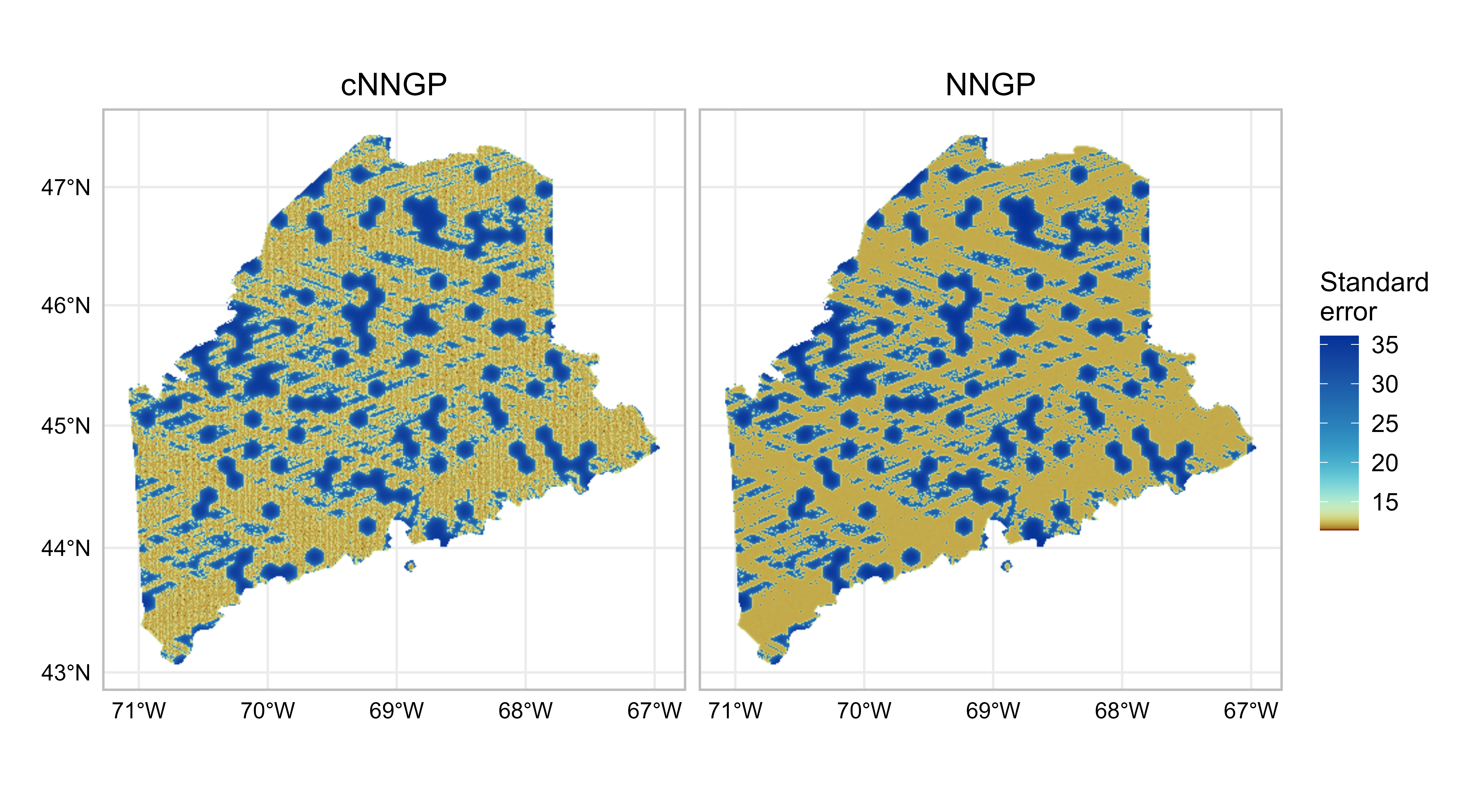}
    \caption{$\,$ Standard error of predicted biomass values.}
    \label{fig:gedi se}
\end{figure}

\begin{figure}
    \centering
    \includegraphics[width=0.9\linewidth]{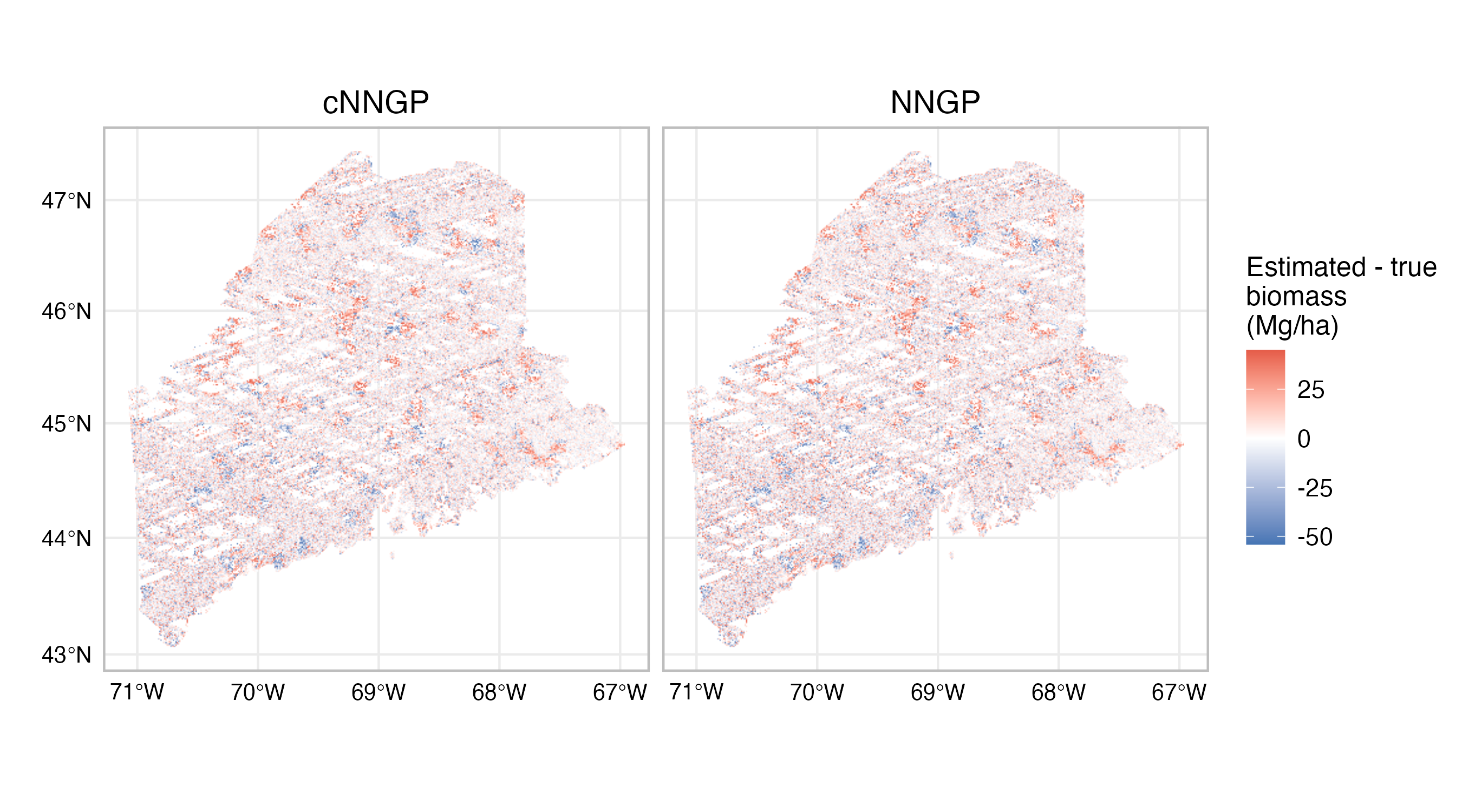}
    \caption{$\,$ Difference between model posterior mean and recorded biomass value. For figure clarity, only the differences within the first and ninety-ninth percentile were included.}
    \label{fig:gedi est-true}
\end{figure}

\begin{figure}
    \centering
    \includegraphics[width=1.1\linewidth]{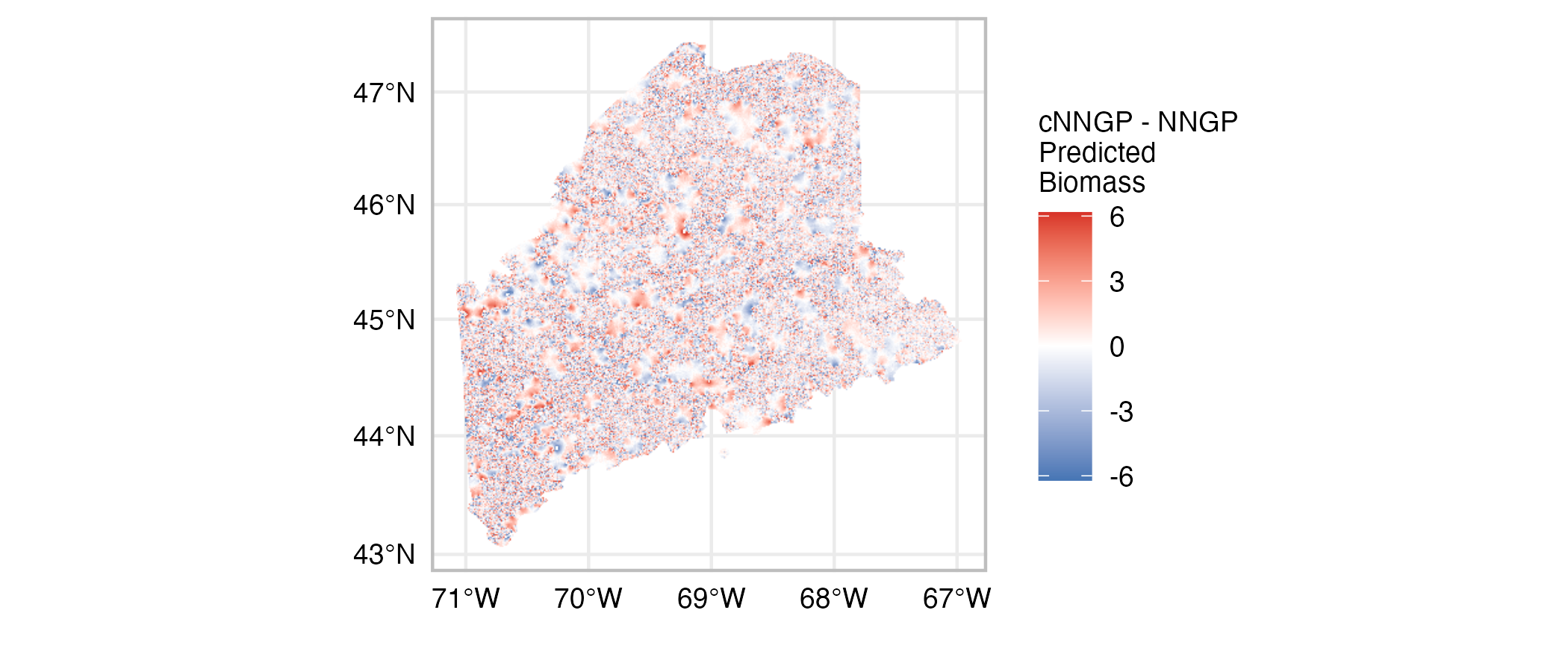}
    \caption{$\,$ Difference between cNNGP and NNGP posterior means. For figure clarity, only the differences within the first and ninety-ninth percentile were included.}
    \label{fig:gedi mean diff}
\end{figure}

\begin{figure}
    \centering
    \includegraphics[width=1.1\linewidth]{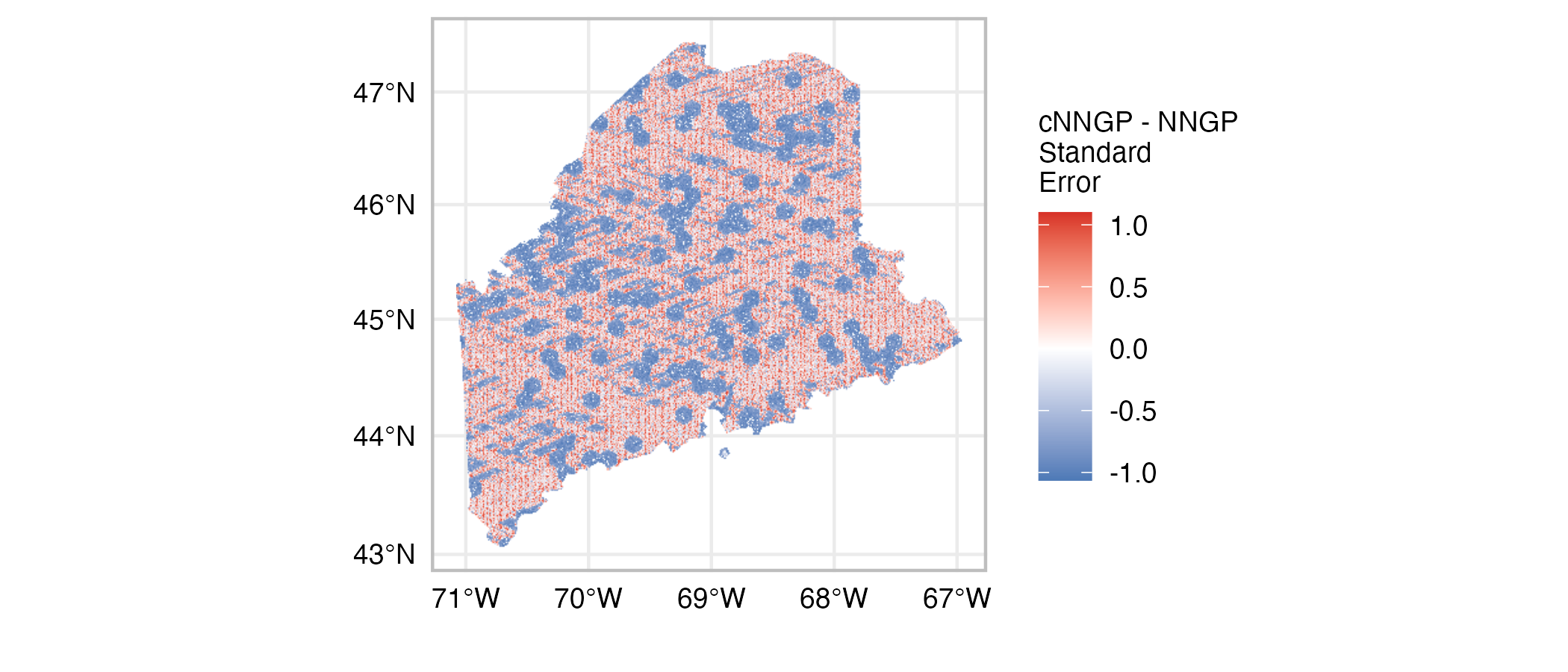}
    \caption{$\,$ Difference between cNNGP and NNGP standard errors. For figure clarity, only the differences within the first and ninety-ninth percentile were included.}
    \label{fig:gedi se diff}
\end{figure}

\end{document}